# Evidence for Large-scale, Rapid Gas Inflows in z~2 Star-forming Disks

R. Genzel[1,2], J.-B. Jolly[1], D. Liu[1], S.H. Price[3], L.L. Lee[1], N.M. Förster Schreiber[1], L.J. Tacconi[1], R. Herrera-Camus[4], C. Barfety[1], A. Burkert[5], Y. Cao[1], R.I. Davies[1], A. Dekel[6], M.M. Lee[7,8], D. Lutz[1], T. Naab[9], R. Neri[10], A. Nestor Shachar[11], S. Pastras[1], C. Pulsoni[1], A. Renzini[12], K. Schuster[10], T.T. Shimizu[1], F. Stanley[10], A. Sternberg[1,11,13] & H. Übler[14,15]

[1]*Max-Planck-Institut für Extraterrestrische Physik (MPE), Giessenbachstraße 1, D-85748 Garching, Germany (genzel@mpe.mpg.de, forster@mpe.mpg.de, linda@mpe.mpg.de)*
[2]*Departments of Physics and Astronomy, University of California, Berkeley, CA 94720, USA*
[3]*Department of Physics and Astronomy and PITT PACC, University of Pittsburgh, Pittsburgh, PA 15260, USA*
[4]*Astronomy Department, Universidad de Concepción, Av. Esteban Iturra s/n Barrio Universitario, Casilla 160, Concepción, Chile*
[5]*Universitäts-Sternwarte Ludwig-Maximilians-Universität (USM), Scheinerstraße 1, D-81679 München, Germany*
[6]*Racah Institute of Physics, The Hebrew University of Jerusalem, Jerusalem 9190401, Israel*
[7]*Cosmic Dawn Center (DAWN), Denmark*
[8]*DTU-Space, Technical University of Denmark, Elektrovej 327, DK-2800 Kgs. Lyngby, Denmark*
[9]*Max-Planck Institute for Astrophysics, Karl-Schwarzschild-Straße 1, D-85748 Garching, Germany*
[10]*Institut de Radioastronomie Millimétrique (IRAM), 300 rue de la Piscine, F-38406, Saint Martin d'Hères, France*
[11]*School of Physics and Astronomy, Tel Aviv University, Tel Aviv 69978, Israel*
[12]*Osservatorio Astronomico di Padova, Vicolo dell'Osservatorio 5, Padova, I-35122, Italy*
[13]*Center for Computational Astrophysics, Flatiron Institute, 162 5th Avenue, New York, NY 10010, USA*
[14]*Kavli Institute for Cosmology, University of Cambridge, Madingley Road, Cambridge, CB3 0HA, UK*
[15]*Cavendish Laboratory, University of Cambridge, 19 JJ Thomson Avenue, Cambridge, CB3 0HE, UK*

## ABSTRACT

We report high-quality Hα/CO, imaging spectroscopy of nine massive (log median stellar mass = 10.65 $M_\odot$), disk galaxies on the star-forming, main sequence (henceforth 'SFGs'), near the peak of cosmic galaxy evolution ($z \sim 1.1 - 2.5$), taken with the ESO-Very Large Telescope, IRAM-NOEMA and Atacama Large Millimeter/submillimeter Array. We fit the major axis position–velocity cuts with beam-convolved, forward models with a bulge, a turbulent rotating disk, and a dark matter (DM) halo. We include priors for stellar and molecular gas masses, optical light effective radii and inclinations, and DM masses from our previous rotation curve analyses of these galaxies. We then subtract the inferred 2D model-galaxy velocity and velocity dispersion maps from those of the observed galaxies. We investigate whether the residual velocity and velocity dispersion maps show indications for radial flows. We also carry out kinemetry, a model-independent tool for detecting radial flows. We find that all nine galaxies exhibit significant non-tangential flows. In six SFGs, the inflow velocities ($v_r \sim 30-90$ km s$^{-1}$, 10%–30% of the rotational component) are along the minor axis of these galaxies. In two cases the inflow appears to be off the minor axis. The magnitudes of the radial motions are in broad agreement with the expectations from analytic models of gravitationally unstable, gas-rich disks. Gravitational torques due to clump and bar formation, or spiral arms, drive gas rapidly inward and result in the formation of central disks and large bulges. If this interpretation is correct, our observations imply that gas is transported into the central regions on ~10 dynamical time scales.

*Unified Astronomy Thesaurus concepts:* High-redshift galaxies (734); Galaxy kinematics (602); Galaxy structure (622); Galaxy evolution (594)

## 1. INTRODUCTION

The cosmic star formation density peaked ~10 Gyr ago ($z \sim 2$, Madau & Dickinson 2014). At that epoch, galaxy halos containing Milky Way mass galaxies first formed in large numbers (e.g., Mo & White 2002). Over the past two decades high-throughput, adaptive optics assisted, near-infrared integral field spectrometers (IFS), such as SINFONI on the ESO-VLT (Very Large Telescope; Eisenhauer et al. 2003; Bonnet et al. 2004), or OSIRIS on the Keck telescope (Larkin et al. 2006), and seeing limited, multiplexed IFSs, such as KMOS at the VLT (Sharples et al. 2012), have become available on 8–10 m telescopes. With these IFSs it has become possible to carry out deep, velocity-resolved (FWHM~80–120 km s$^{-1}$) spectroscopic imaging of Hα in the $z \sim 0.6 - 2.6$ main sequence (MS; Whitaker et al. 2012; Speagle et al. 2014; Whitaker et al. 2014) star-forming galaxies (SFGs). At around the same time, subarcsecond millimeter interferometric imaging of CO rotational lines has become feasible in the same redshift range with the sensitive IRAM-NOEMA and Atacama Large Millimeter/submillimeter Array (ALMA) arrays.

Over the past decade, we have undertaken two main IFS surveys of high-$z$ galaxy kinematics. At the ESO-VLT, we carried out **SINS** and **zC-SINF** with SINFONI (Förster Schreiber et al. 2006; Genzel et al. 2006; Förster Schreiber et al. 2009, 2018; Förster Schreiber & Wuyts 2020), and **KMOS**[3D] with KMOS (Wisnioski et al. 2015, 2019) in about 750 $z \sim 0.6 - 2.6$ SFGs covering the mass range of $\log(M_*/M_\odot) = 9.5-11.5$. At IRAM-NOEMA we observed the CO 3–2/4–3 lines in about 200 SFGs in the same redshift and mass range with the IRAM-NOEMA millimeter interferometer as part of the **PHIBSS 1** and **2,** and **NOEMA³ᴰ** surveys (Tacconi et al. 2010, 2013, 2018, 2020). In total we have assembled ~1000 IFS data sets of MS SFGs. In our SINS, KMOS³ᴰ, and NOEMA³ᴰ surveys, we have emphasized deep integrations, for high-quality data on individual galaxies. The highest quality data are collected as part of the **RC100 sample** (100 galaxies between $z = 0.6$ and 2.5, Nestor et al. 2023, see also Genzel et al. 2020, Price et al. 2021).

These (and other) studies have established that 60%–80% of the more massive, near-MS SFGs at $z \sim 1 - 2.5$ **are rotationally supported disks, not major mergers as expected from earlier work** (see Erb et al. 2004; Förster Schreiber et al. 2006; Genzel et al. 2006; Kassin et al. 2007; Förster Schreiber et al. 2009; Kassin et al. 2012; Swinbank et al. 2012; Wisnioski et al. 2015; Stott et al. 2016; Simons et al. 2017; Swinbank et al. 2017; Wisnioski et al. 2019). The disks are turbulent and geometrically thick with $v_{\rm rot}(R_{\rm e})/\sigma_0 \sim 3-10$ (Wisnioski et al. 2015; Simons et al. 2017; Wisnioski et al. 2019; Übler et al. 2019). Here $v_{\rm rot}(R_{\rm e}) = v_c$ is the inclination and beam smearing corrected, intrinsic rotation velocity of the disk at the half-light radius $R_{\rm e}$, and $\sigma_0$ is the average velocity dispersion of the (outer parts of the) disk, after removal of beam-smeared rotation and instrumental line broadening. These conclusions initially surprised many in the field, given the framework of the growth of dark matter (DM) haloes by merging of smaller predecessors in the Lambda cold DM cosmology (e.g., Frenk et al. 1985). Over the last decade, it has become clear that major mergers are relatively rare ($\sim(3~{\rm Gyr})^{-1}$) for massive galaxies (Fakhouri & Ma 2008; Neistein & Dekel 2008; Genel et al. 2009). Most of the growth of massive galaxies at $z \sim 1 - 3$ occurs through accretion of diffuse gas from the intergalactic medium (IGM) and circumgalactic medium (CGM), as well as through minor mergers, followed by settling of the gas in rotating disks (Mo et al. 1998; Dekel et al. 2013). Internal star formation in the gas-rich, interstellar medium then leads to the growth of the stellar component, while stellar and active galactic nuclei (AGN) feedback eject some of the gas back into the CGM (the **cosmic 'baryon cycle'**: Dekel et al. 2009ab; Bouché et al. 2010; Guo et al. 2010; Steidel et al. 2010; Lilly et al. 2013; Bower et al. 2017; Förster Schreiber & Wuyts 2020; Péroux & Howk 2020; Tacconi et al. 2020).

As mentioned above, typical velocity dispersions in the ionized and molecular interstellar medium of $z \sim 2$ SFGs are about 2 times greater than those in the local Universe (ranging from $\sim 15-30$ km s$^{-1}$ at $z \sim 0$ to $\sim 40-70$ km s$^{-1}$ at $z \sim 2$, see Förster Schreiber et al. 2006; Kassin et al. 2007; Law et al. 2007; Förster Schreiber et al. 2009; Kassin et al. 2012; Jones et al. 2013; Wisnioski et al. 2015; Simons et al. 2017; Turner et al. 2017; Förster Schreiber et al. 2018; Johnson et al. 2018; Übler et al. 2019; Wisnioski et al. 2019). Some compact SFGs are even 'dispersion dominated' ($v_c/\sigma_0 \leq 1$, Newman et al. 2013). However, the scatter of $v_c/\sigma_0$ is large at a given redshift (Übler et al. 2019). Some systems appear to be much colder (Di Teodoro et al. 2016; Rizzo et al. 2020; Fraternali et al. 2021; Rizzo et al. 2021; Lelli et al. 2023), either because of being special cases (e.g., dusty lensed systems), or because of differences in the data analysis.

## 2. NONCIRCULAR MOTIONS

### 2.1. Sample Selection and Analysis

The topic of this paper is the evidence for noncircular motions in $z \sim 1 - 2.5$ SFGs. To investigate whether large radial motions occur in isolated, gas-rich high-$z$ SFGs, the challenge is to separate such second-order, streaming motions, from the dominant first-order rotation. The goal of this paper is to carry out such a study for a sample of SFGs, with a range in redshift and mass. This goal requires outstanding data quality and resolution. Our parent sample is RC41/RC100 (Nestor et al. 2023, see also Genzel et al. 2020; Price et al. 2021), and we refer to the extensive discussions in these papers for sample details. RC100 covers quite well the near-MS, massive ($\log(M_{\rm baryon}/M_\odot) > 9.3 - 11.4$) SFG population, at and around the peak of cosmic star/galaxy formation, $z = 0.6-2.5$. Almost all RC100 galaxies have Hubble Space Telescope (HST) imaging coverage, as well as H$\alpha$ imaging spectroscopy from SINFONI and KMOS. For 10 SFGs there is NOEMA or ALMA imaging spectroscopy of CO 3–2 or 4–3 emission. We adopt standard cosmological parameters $H_0 = 70$ km s$^{-1}$ Mpc$^{-1}$, $\Omega_m = 0.3$, $\Omega_\Lambda = 0.7$.

Starting with the full RC100 sample we selected initially about two dozen well-resolved (resolution FWHM $\leq 0.5''$), large ($R_{\rm e} > 4$ kpc,) galaxies with deep integrations ($> 10$ hr), and inclination $< 70°$, such that there are enough independent spaxels justifying a 2D analysis of the data. We eliminated mergers, galaxies with massive companions, and strong AGN (because of very broad emission lines, with the exception of D3a_15504 where the AGN only affects the central spaxels and the data are of excellent resolution). This left 11 candidates. Of these, three did not have sufficient spatial coverage for a reliable investigation of the 2D kinematics. One galaxy was a very face-on, clumpy ring, again not suitable for investigating 2D kinematics. This leaves seven candidates passing all criteria. In the end, we added one smaller ($R_{\rm e} = 2.6$ kpc), low inclination system with excellent high-resolution data (zC_403741, $z = 1.446$). We also added the $z = 1.1$ galaxy EGS_13035123 (resolution $0.5''$), which is very large ($R_{\rm e} = 10$ kpc) and has exquisite CO 3–2 and 4–3 NOEMA data. The salient parameters of the final sample of nine SFGs are given in Table 1. We carried out several independent **forward modeling analyses** of these galaxies. In addition, we analyzed the data independently with **kinemetry**.

## 2.2. 2D & 3D Analyses of the Velocity Fields

Qualitative evidence for large-scale inflow/outflow signatures in rotating disks is well known to occur mainly along or near the minor axis **if these flows are axisymmetric** (van der Kruit & Allen 1978). In these cases, the projected velocity $v_{obs}$ at a given galaxy radius contains sinusoidal components in the angle $\theta$ in the plane of the galaxy (relative to the major axis, positive east of north; Figure 1 in van der Kruit & Allen 1978),

$$v_{obs} = v_{sys} + v_\theta \times \sin i \times \cos \theta + v_R \times \sin i \times \sin \theta,$$

$$sky\ (r,\phi):\ \tan \theta = \tan(\phi - \phi_0) / \cos i,$$

$$galaxy\ (R, \theta):\ R = r \times \cos(\phi - \phi_0) / \cos \theta \quad (1),$$

where $i$ is the inclination of the galaxy plane relative to the sky plane ($i = 0$ or $180°$ is face-on), $\phi$ is the angle on the sky (positive east of north), and $\phi_0$ is the angle on the sky of the major axis of the galaxy (see Figure 1 for an illustration of these angles). Alternatively, radial streaming motions can also occur **off the minor axis if there is a non-axisymmetric perturbation of the gravitational field**, such as an internal stellar or gas bar (Roberts, Huntley & van Albada 1979; van Albada & Roberts 1981; Athanassoula 1992; Bournaud & Combes 2002; Binney & Tremaine 2008). Another cause for radial flows can be a substantial, external perturbing mass (Toomre & Toomre 1972, Binney & Tremaine 2008), or gas flows from the CGM with an angular momentum different from that of the central disk (A. Dekel, private communication).

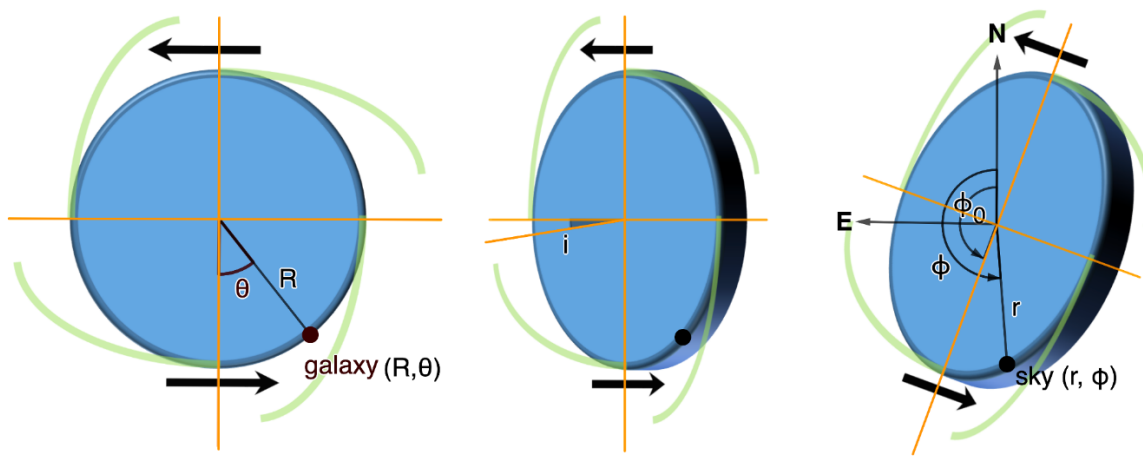

*Figure 1: Angles in equation (1). Left: the azimuthal angle $\theta$ within the galactic plane. Middle: the inclination angle $i$ of the galaxy plane relative to the sky plane (face-on is $i = 0$ (counterclockwise) or $180°$ (clockwise)). Right: the position angle of a galactic position projected on the sky, $\phi$ (positive east of north), and the position angle of the major axis of the galaxy projected on the sky, $\phi_0$.*

## 2.3. Analysis

### 2.3.1. Kinemetry

In a **kinemetry** analysis of **motions in a planar disk** (Krajnović et al. 2006; Shapiro et al. 2008) the data at each radius $R$ (projected to an ellipse on the sky) are fit with an expansion in $\sin \theta$ and $\cos \theta$,

$$K(\theta) = A_0 + A_1 \times \sin \theta + B_1 \times \cos \theta$$
$$+ A_2 \sin(2\theta) + B_2 \cos(2\theta) + \ldots \quad (2).$$

Here the coefficient $B_1$ is a measure of pure tangential rotation at $R$, while $A_1$, $A_2$ etc. measure the amplitude of radial velocities. We performed the **kinemetry** analyses up to second order; higher-order decomposition did not improve nor change the results significantly. In the end, we used the dominant lowest order, $A_1 \sim v_r$. Averaging over independent annuli in $R_j$ yields a quantitative measure for the relative magnitude of radial to tangential motions,

$$<v_r / v_c> = \left[ \sum_{j(R)} \frac{(|A_1| + \ldots)_j}{(|B_1|)_j} \right] / \sum_{j(R)} j \quad (3).$$

Note that in equations (2) and (3) the major axis of the galaxy is fixed at $= 0$ (and $\phi_0$ on the sky). In a purely rotating system, this angle is identical to the angle of the kinematic axis, or *line of nodes* (i.e., the angle defined by the direction between the largest and smallest velocities). This is not the case for a galaxy with substantial radial velocity. In that case equation (1) shows that the kinematic axis is rotated relative to the structural major axis (for instance the major axis of the ellipse that best fits the $H$-band surface brightness distribution) by an angle $\theta_0$,

$$\theta_0 = -\arcsin(v_r / v_t),$$
$$\text{where } v_t = \sqrt{(v_c^2 + v_r^2)} \quad (4).$$

For instance, for $v_r/v_t \sim 0.2$–$0.3$ this angle is $\theta_0 = 11 - 17°$, which is detectable in our highest-quality data sets.

### 2.3.2. DYSMAL Forward Modeling

We use forward modeling to infer the baryonic mass distribution, DM content and bulge-to-total ratios mainly from the near-rotation axis, 1D cuts in velocity, velocity dispersion, and light, with additional priors on the mass center, inclination, and effective radii mainly from rest-frame optical HST imaging (Genzel et al. 2017,2020; Price et al. 2021; Nestor et al 2023). As in our earlier work (Genzel et al. 2006; Burkert et al. 2016; Wuyts et al. 2016; Genzel et al. 2017; Lang et al. 2017; Übler et al. 2017, 2018) we use a parameterized, input mass distribution to establish the best fit models for a given Hα or CO data set. This mass model is axisymmetric and is the sum of an unresolved bulge, a rotating flat disk of Sérsic index $n_s$, effective (half-light) radius $R_e$, (constant) isotropic velocity dispersion $\sigma_0$, and a surrounding halo of DM. As discussed in more detail in the above references, we compute from this mass model 3D data cubes ($I(x, y, v_z)$) of the disk gas, convolved with a 3D kernel describing the instrumental point-spread function PSF ($\delta x, \delta y, \delta v_z$) of our measurements. We then compare directly this *beam smeared* model to the observed data, and vary model parameters to obtain the best fits. The most common approach, adopted in this paper, is to extract velocity centroids (from Gaussian fits) and velocity dispersion cuts from a suitable software slit along the dynamical major axis of the galaxy, for both the model and measurement cubes. We typically use a constant software slit width (typically ~1–1.5 × FWHM of the PSF).

We use the analysis tool **DYSMAL** (Genzel et al. 2006; Davies et al. 2011; Genzel et al. 2017, 2020; Price et al. 2021; Liu et al. 2023; Nestor et al. 2023). **For a detailed description of this tool, we refer the reader to Section 2.1 and Appendix A of Genzel et al. (2020) and Section 3 and the Appendix of Price et al. (2021).** Price et al. (2021) have shown that most of the information on the intrinsic rotation curve, and thus on the mass distribution, is contained in these 1D cuts (see also Genzel et al. 2006, 2008). For most high-$z$ galaxies the number of independent spaxels in the 3D data cubes is 40–200 (8–15 spatial resolution elements, times 5–15 spectral resolution elements) so that the additional information on the mass model from off-axis spaxels is modest, especially for more edge-on systems (the 1D cuts involve typically 30–100 of these spaxels). In comparison to nearby galaxies, the *spider diagram* off the main axis is more sensitive to inclination and radial streaming than to the intrinsic mass distribution (e.g., van der Kruit and Allen 1978; Price et al. 2021).

With the fit results from **DYSMAL** in hand, we next compute a 3D model data cube, with the same resolution as the observed data cubes of the galaxy. For constructing residual cubes (data minus model) that are sensitive to non-tangential motions, the galaxy's center location and the orientation of the major axis on the sky are critical parameters. We use priors from the continuum mapping (with HST, and also (sub)-millimeter continuum if available), or from the integrated line intensity distribution(s) to determine the center on the sky. This is especially trustworthy if there is a massive central bulge. The velocity dispersion distribution derived from the line(s) is another reliable indicator, since the beam-smeared velocity dispersion naturally has a maximum at the mass center of a rotating galaxy. The remaining parameter is the angle of the rotation axis on the sky. Following the discussion above (Equation (4)), the angle of the line of nodes is not an accurate measure of the galaxy's major axis if radial streaming velocities are large. Instead, we place stronger weight on the orientation of the major axis in the continuum light distribution (typically *H*-band HST), or in the integrated line intensity distribution(s).

### 2.4. Mock Galaxies

To demonstrate more clearly the signatures of radial streaming motions we have constructed mock galaxy models. We have taken a model rotating galaxy, with parameters like Q2343_BX610 (Section 3.1, Table 1). We constructed the best-fit data cubes for three cases: (1) no radial motion ($v_c(R_e = 4 \text{ kpc}) = 295 \text{ km s}^{-1}$, inclination = 39°, position angle of the major axis at $PA_\text{major} = -25 \pm 8°$), (2) axisymmetric inflow with $|v_r| = 100 \text{ km s}^{-1}$, and (3) an inflow along an axis with $PA = -156°$ ($\delta PA = PA_\text{inflow} - PA_\text{major} = -131°$). The latter example mimics the inflow in a barred galaxy. Figure 2 shows the resulting velocity fields (blue approaching, red receding, green systemic velocity) for the pure tangential motion case (left), and the axisymmetric (top center) and bar (bottom center) flows. As expected, for the two cases with radial motions the axis symmetry of the velocity field is broken, and the isovelocity contours exhibit a **characteristic S-shape** (see the detailed discussion and local Universe examples in van der Kruit & Allen 1978, and references therein). The radial velocity structure becomes more obvious in the residual maps after subtraction of the intrinsic rotational component of the velocity field (top and bottom right). A characteristic property of an axisymmetric radial streaming is the cone-shaped residuals along the minor axis of rotation (top right). Streaming along a well-defined axis has a sharper signature along the streaming axis (bottom right).

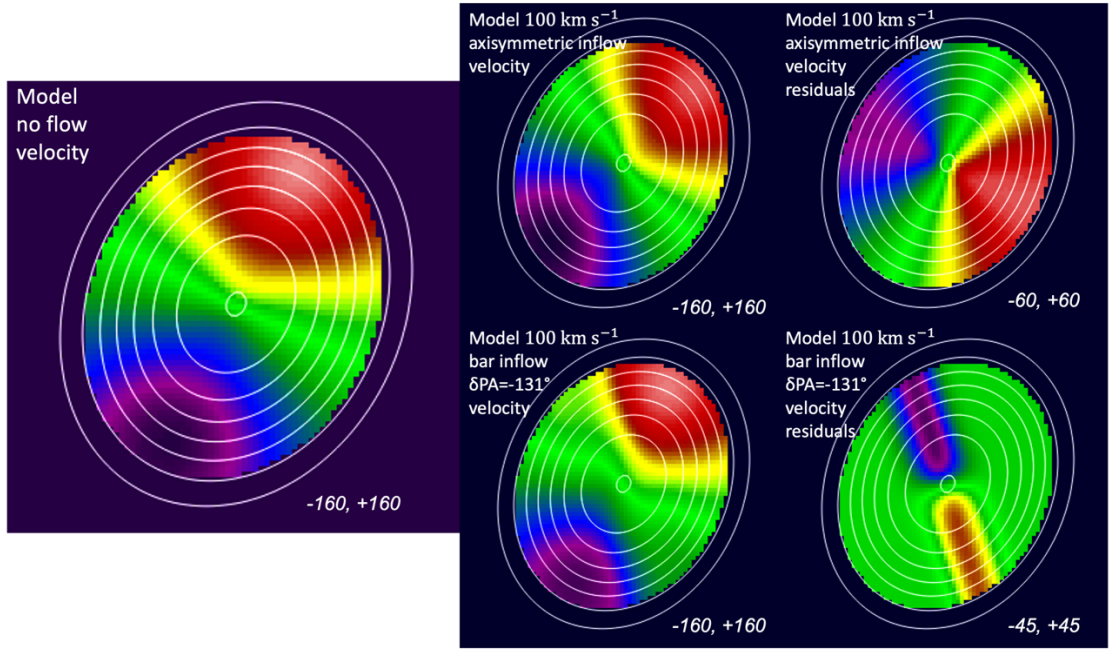

*Figure 2. Model velocity maps of a rotating disk galaxy with parameters like Q2343_BX610 ($v_c(R_e = 4$ kpc$) = 295$ km s$^{-1}$, $i = 39°$, $PA_{major} = -25°$). Left: Model velocity map without any radial motion. Top center and right: Model velocity (center) and residual ($v_c(no\ flow) - v_c(flow)$) velocity map of a galaxy with 100 km s$^{-1}$ axisymmetric inflow. Bottom center and right: Same for a 100 km s$^{-1}$ inflow along a bar at $PA_{inflow} = -156°$, or $\delta PA = PA_{inflow} - PA_{major} = -131°$ with respect to the kinematic major axis of rotation.*

### 2.4.1 Converging or Diverging Radial Flows?

How can one distinguish radially converging flows (**inflows**) from radially diverging flows (**outflows**)? Consider a symmetrically rotating disk at an inclination of 130° (Figure 3) and $PA_{major} = 180°$ such that the maximum approaching velocity is at the top and the rotation on the sky is clockwise. Inflows and outflows are distinguishable by the sign of their projected velocity relative to the rotational motion at $PA_{major} = 180°$. Then the characteristic of inflow (outflow) is that in quadrants 1 and 2 the projected radial velocity has the opposite (same) sign as the rotation velocity at $PA = 0°$, while in quadrants 3 and 4 the projected radial velocity has the same (opposite) sign of that at $PA = 0°$. We derived the orientation of the rotational motion on the sky (clockwise or counterclockwise) using the spiral arms when observed (by assuming trailing arms, see Toomre 1981), or using attenuation/color criteria otherwise (see Appendix A for individual cases).

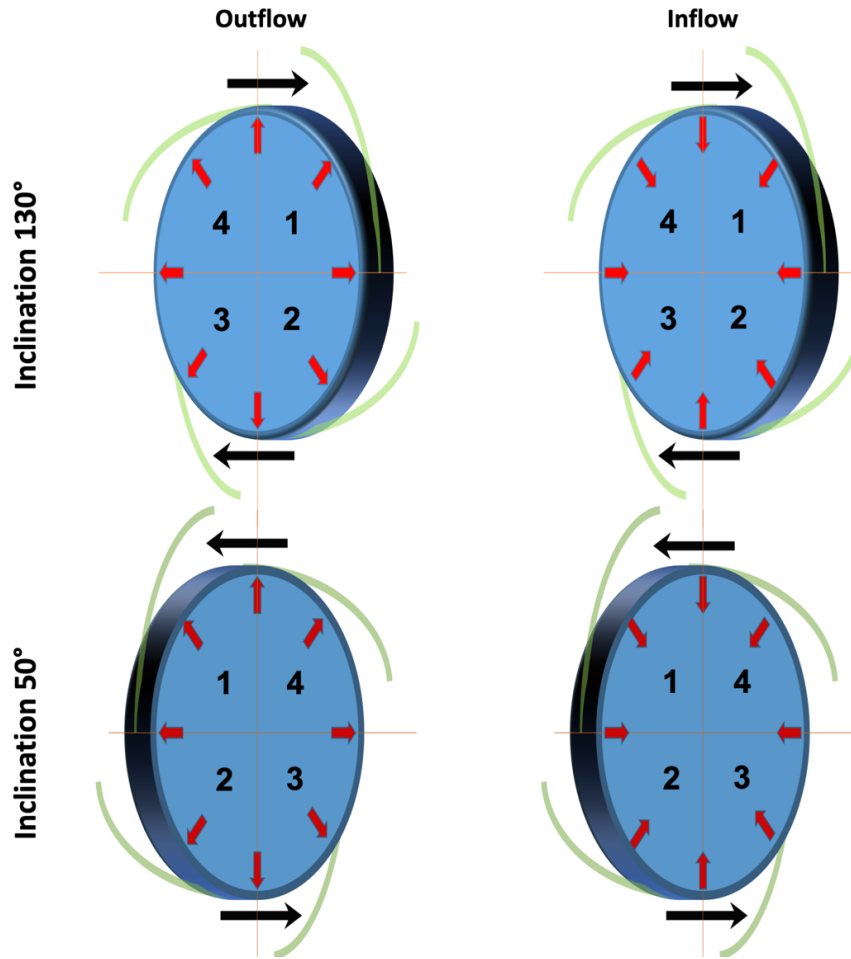

*Figure 3.* Cartoons of axisymmetric, in-plane outflows (left) and inflows (right) in a galaxy viewed at an inclination of $130°$ (top row; rotating clockwise) and $50°$ (bottom row, rotating counterclockwise), respectively. The position angle of the major axis is $PA_{\mathrm{major}} = 180°$ in all panels. The largest negative (approaching) and positive (receding) velocities thus occur near $PA = 0°$ and $180°$. In all cartoons, we also added trailing spiral arms. Labeled quadrants 1 to 4 are along the in-plane rotation direction from the approaching side. Then the characteristic of inflow (outflow) is that, for the inclination $> 90°$ case (top row), in quadrants 1 and 2 the projected radial velocity has the opposite (same) sign from the projected rotation velocity at $PA = 0°$, while in quadrants 3 and 4 the projected radial velocity has the same (opposite) sign of the projected rotation velocity at $PA = 0°$. If the inclination of the galaxy is less than $90°$ (bottom row), these relative signs stay the same but the rotation is now counterclockwise.

# 3. RESULTS

## 3.1. Analysis of Q2343_BX610

We first present the extensive Hα, CO, and submillimeter-dust continuum data sets of the $z = 2.2$ galaxy Q2343_BX610 (see also Erb et al. 2004; Steidel et al. 2004; Förster-Schreiber et al. 2006; Genzel et al. 2008; Förster-Schreiber et al. 2009, 2018; R. Herrera-Camus et al. in preparation). Figure 4 (left panel) shows the combination of HST $J$ and $H$ images (blue, and green), with the ALMA-integrated CO 4–3 image (red), all at ~0.2″ resolution. The right panel of Figure 4 shows the combination of HST $J+H$ (blue), with integrated AO Hα (K100, green), and the ALMA 150 GHz continuum (rest-frame 480 GHz, or 630 μm, red). The spiral arms are clearly identified, allowing us to derive the direction of the rotational motion on the sky (counterclockwise). At the bottom we show cuts along the major axis of integrated line intensity, velocity, and velocity dispersion for three data sets: 0.28″ Hα data with adaptive optics (AO or K100) mode (integration time 8 hr, blue open circles), 0.5" Hα data with AO plus seeing limited (non-AO or K250) modes (integration 34 hr, filled blue circles), and 0.2″ CO 4–3 data (integration 24 hr, crossed red squares). Figure 5 shows the derived 2D velocity and velocity dispersion maps in CO 4-3, Hα (non-AO and AO), Hα (AO only), and [NII] 6583 (non-AO and AO). Figure 6 displays the velocity and velocity dispersion residual maps, after subtracting the **DYSMAL** model maps from the data, as described in Section 2. Finally, Figure 7 shows the **kinemetry** analysis of the highest resolution line data (CO 4–3). Tables 1– 3 summarize the salient intrinsic parameters of the galaxy, including molecular gas fraction inferred from scaling relations (Tacconi et al. 2018, 2020) and CO 4–3 flux, the Toomre parameter inferred from equation (10), the derived inflow velocity (assumed to be in the rotation plane of the galaxy), and the relevant angles and morphologies. In Table 2 we also compare the empirically resolved results to the analytic inflow velocities expected in gas-rich, Toomre unstable disks, with large expected inflow rates due to gravitational torques, viscous angular momentum transport, and dynamical friction (Equation (11) with a~1.75, $\zeta$=2, and $\gamma$=0).

Our findings in Q2343_BX610 can be summarized as follows,

1. Q2343_BX610 is a typical, massive ( $\log(M_{baryon}/M_\odot) \sim 11$, $v_c \sim 300$ km s$^{-1}$ ), gas-rich ( $f_{gas} \sim 0.62$ ), turbulent ( $\sigma_0 = 55$–65 km s$^{-1}$ ) and large ( $R_e \sim 4$ kpc) main-sequence SFG at $z = 2.2$. Its optical stellar bulge (uncorrected for extinction) is modest (bulge-to-total ratio ($B/T$)~0.12) but prominent in molecular gas and dust (Figure 4). Its inclination is low (39°), allowing a detailed mapping of the near minor axis kinematics and structure.

2. Q2343_BX610 is somewhat unusual compared to other $z\sim 2$ SFGs in RC100 (Nestor et al. 2023) in that it appears to have a prominent bar-like structure in the HST $J+H$ continuum (i.e., rest-frame $V$ band), in Hα, and less prominently also in CO 4–3. This *bar* is along $PA \sim 22$–25°, offset by ~49° on the sky, and ~56° in the plane of the galaxy, away from the kinematic major axis of the galaxy. At the tips of this structure Hα emission and $J$-band continuum are very bright, reminiscent of local barred galaxies (e.g., Combes et al. 2014).

3. Considering the differences in resolution and angular sizes (CO is the most compact, while Hα and [NII] 6583 are more extended, especially the deepest Hα with the lowest resolution), the derived 1D and 2D velocity and velocity dispersion distributions in Figure 4 (bottom) and Figure 5 are in remarkable agreement. Note that the velocity dispersion in CO 4–3 and Hα with AO is less centrally peaked and lower in amplitude than the respective combined AO and seeing limited (K100+K250) Hα and [NII] 6583 distributions. This is mainly caused by beam smearing of the velocity distribution and less due to intrinsic velocity dispersion. The CO map has higher angular resolution than the optical tracers, and thus suffers less beam smearing.

4. The major rotation axis derived from the Hα and [NII] velocity maps differs slightly from the one extracted using CO. The CO major axis lies along $PA \sim -33°$, while the Hα and [NII] rotation axes are at a lower angle, of $-18°$ to $-25°$. We have investigated two options, one with a common PA for all tracers, $PA \sim -25 \pm 8°$, and one with different PAs for each tracer as suggested by the data. The first option leads to very different residuals in the velocity and velocity dispersion maps of all three tracers. The second option, however, yields comparable residual patterns within the uncertainties. For this reason, we prefer the second option, which is shown in Figures 5 and 6. As a consequence, the PA of the rotation axis of the more compact and dense molecular gas differs from the more diffuse, outer-disk ionized gas by about 10° to 15°.

5. The CO 4–3 and Hα high-resolution, isovelocity contours at low absolute velocity (green color in Figure 5) are along $PA \sim -156°$, approximately identical to the near-IR bar. The minor axis is at $PA_{minor} = 65°$, ~40° off the bar.

6. The velocity and velocity dispersion residuals in Figure 6 (after subtracting the best fitting **DYSMAL** model maps) also compare reasonably to very well in the four maps. In all velocity residual maps, there is a linear, fairly narrow blue-red velocity gradient ridge, centered on the nuclear bulge position. The position angle $PA_{resid}$ and the half amplitude velocity gradient $v_{resid}$ in the four maps are: $-153$ (°), 72 (km s$^{-1}$) (CO), $-164$, 57 ([NII] 6583), $-155$, 45 (Hα AO) and $-157$, 35 (Hα non-AO+AO). Taking a weighted average we find $PA_{resid} = -156°$ and $v_{resid} = 60$ km s$^{-1}$ (sky projected). For comparison, two independent additional analyses in our team yield half amplitudes of 50 and 70 km s$^{-1}$. Given the uncertainties in zero-points and S/Ns of the maps, the typical uncertainty of the position angle is $\pm 3°$–5°, and the uncertainty in the amplitude $\pm 12$ km s$^{-1}$.

Note that the residual velocity dispersion maps all exhibit a minimum at the center, surrounded by a shallow ring of positive residuals.

7. We have shown in Section 2.4 that these properties are expected for a converging flow (inflow) along the stellar/gas bar in Q2343_BX610 (and a null-hypothesis test described in Appendix B also supports this). The comparison is less favorable for an axisymmetric inflow, as this would result in a broader cone of residuals along the minor axis ($PA_{\mathrm{minor}} = +65°$). If the flow is in the plane of the galaxy, its intrinsic half amplitude is $v_r = v_{\mathrm{resid}}/\sin i = 95 \pm 24 \text{ km s}^{-1}$. The ratio of inflow to tangential velocity is $v_r/v_c = 0.32$. The **kinemetry** analysis of Q2343_BX610 (Figure 7) clearly shows the highly significant presence of a constant $A_1(\sin\theta)$ component with $|A_1|/B_1 = v_r/v_c = 0.34$, in excellent agreement with the **DYSMAL** fitting result.

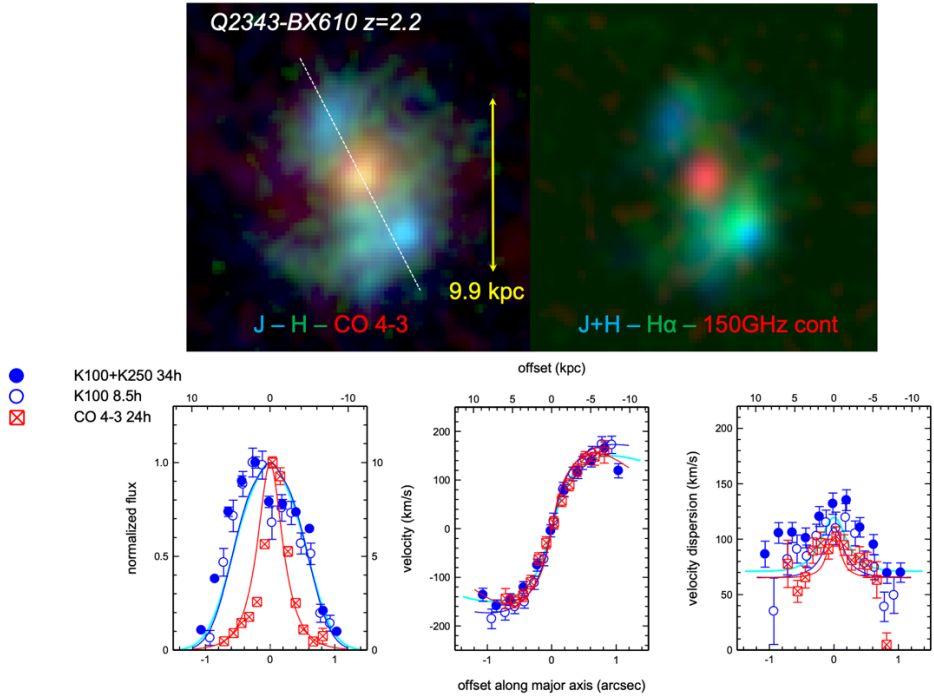

*Figure 4. Top left: Combined HST J+H continuum (blue and green, corresponding to the rest-frame V band), and integrated ALMA CO 4–3 (red), all at ~0.2″ FWHM resolution. The dotted white line marks the direction of the bar-like structure. Top right: combined HST J+H continuum (blue), and integrated ALMA rest frame 460 GHz continuum (red), and Hα-integrated line (green), at ~0.25″–0.3″ FWHM resolution. Bottom: 1D cuts along the major axis at $PA_{\mathrm{major}} \sim -20$ to $-30°$ in CO 4–3 (red squares), Hα AO (open blue circles), and Hα AO + non-AO (filled blue circles) in line intensity (left), line velocity (center) and line velocity dispersion (right).*

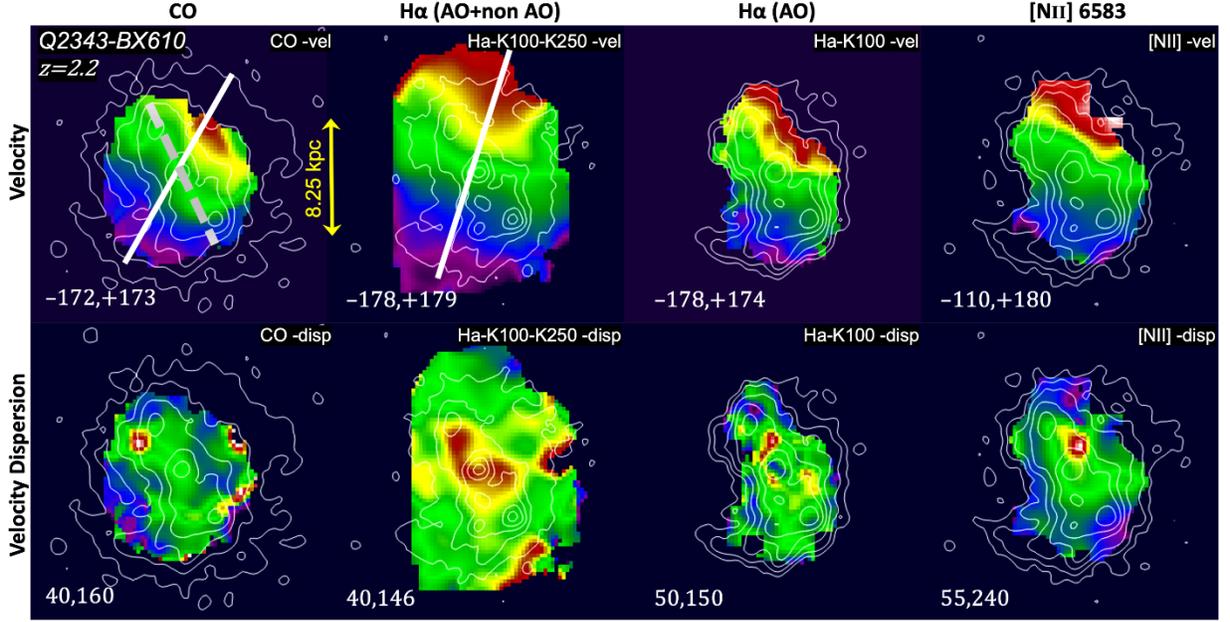

*Figure 5. Velocity (top) and velocity dispersion (bottom) maps in CO 4–3 (left), Hα AO + non-AO (second from left), Hα AO (third from left) and [NII] 6583 (right), superposed on the HST H-band map in contours. The color indicates the velocity amplitude with the numbers at the bottom of each panel giving the minimum (purple) and maximum (red) values. The yellow arrow denotes 1". The kinematic major axis of the galaxy at $PA_{major} \sim -25° \pm 8°$ ($-33$ for the most compact CO and $-18$ for the most extended Hα) is marked as a white line, and the zero isovelocity contour ridge (green color) at $PA \sim 24°$ is indicated by the gray dashed line in the first panel.*

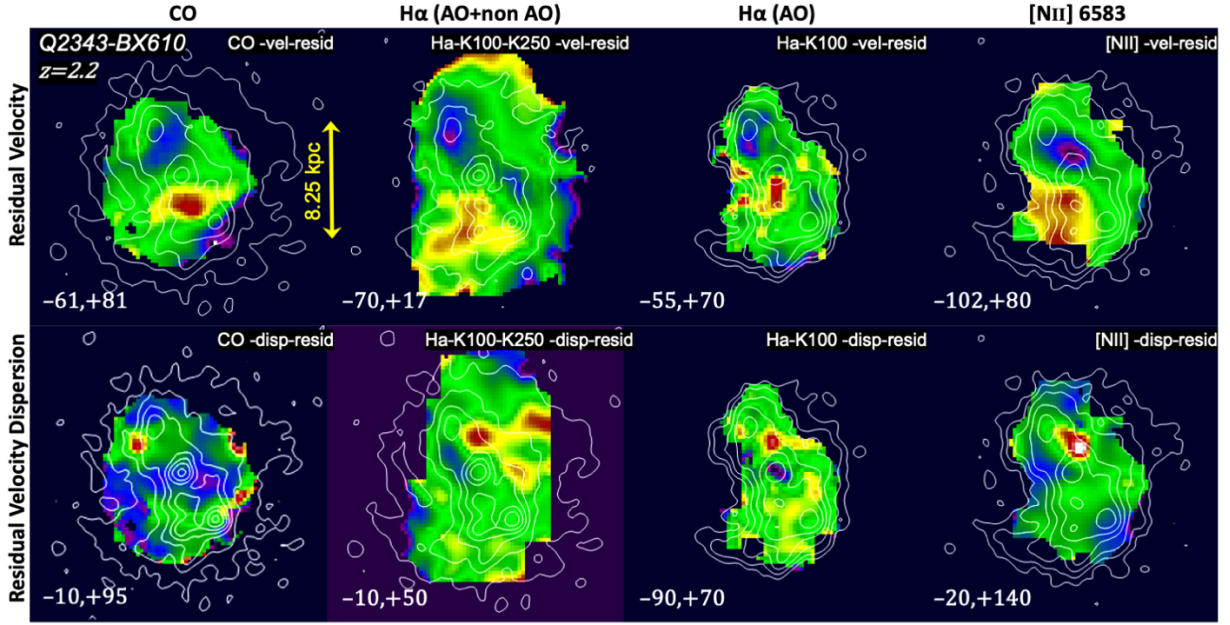

*Figure 6. Velocity residual (top) and velocity dispersion residual (bottom) maps of the CO 4–3 (left), Hα AO + non-AO (second from left), Hα AO (third from left), and [NII] 6583 AO + non-AO data (right), after subtracting the corresponding **DYSMAL** model maps from our data. Color scheme and amplitudes are the same as in Figure 5.*

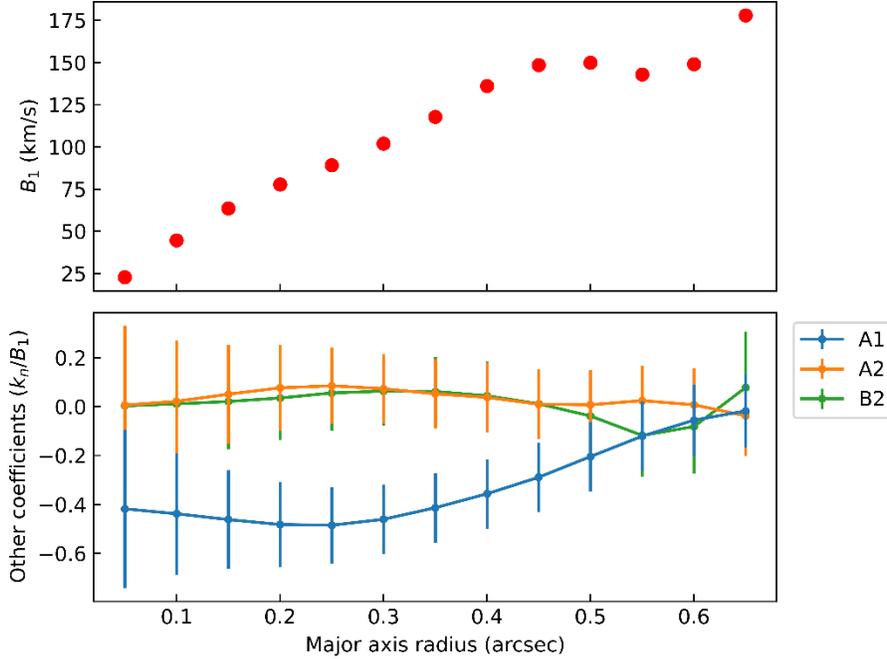

*Figure 7. Kinemetry analysis of the highest resolution line data of Q2343_BX610, the 0.2" CO 4-3 data from ALMA (see equations 2-4). Top: Rotation parameter $B_1$ as a function of galactocentric radius R, extracted for ellipses of $q = 0.75$ along $PA = -25°$. Bottom: Ratio of $A_1/B_1$, $A_2/B_1$, and $B_2/B_1$ as a function of distance from the center. The dominant non-tangential kinemetry parameter is $A_1$ with $A_1/B_1 = -0.2 \cdots -0.44$. Note that this in excellent agreement with the $v_r/v_c(R_e) = -0.32$ from the **DYSMAL** analysis in this line (Table 2).*

In summary, we have analyzed the ~2 kpc scale gas kinematics of the massive z=2.2 SFG Q2343_BX610 in four different tracers. Our results on this galaxy are summarized in Figures 4–7 and Tables 1–3. From forward modeling with **DYSMAL**, as well as **kinemetry** of these 2D velocity and velocity dispersion maps all data sets agree very well to second order (velocity residuals). **Assuming that the gas motions occur in a plane, the gas kinematics in this gas-rich rotating disk galaxy exhibits a rapid (~95 km s$^{-1}$) converging flow (inflow), at about a third of the rotation speed**. We do not see a corresponding outflow (as would be expected in galaxies with bar or spiral arm streaming, see Section 4.1). Gas near the bar at $R_e$ will be transported to the central regions in 2.5 dynamical, or 0.4 orbital times, suggestive of efficient *compaction* (Zolotov et al. 2015, Dekel & Burkert 2014). The prominent central concentration of CO 4–3 and 630 μm (rest frame) continuum compared to Hα and stellar light is consistent with such rapid gas transport to the nucleus. The lack of an outflowing component could perhaps be caused by our data tracing mainly dense gas (the outflowing gas in bar flows is predicted to have lower density than the inflowing component, see Roberts, Huntley and van Albada 1978). **Because the flow is not axisymmetric but along a single direction (the bar), the total rate of inflow is likely a (modest) fraction β of the maximum value, $dM/dt = β \times M_{gas}(R_e) \times v_r/R_e$, with $β \sim 0.2$** (Equation (12); and thus $β\, v_r$ indicates the azimuthally-averaged inflow velocity).

### 3.2. Analysis of Eight Additional SFGs

We analyze eight additional galaxies in the same way as for Q2343_BX610, based on our RC41/RC100 papers. In the Appendix A, we present all their analysis figures. Table 1 summarizes the salient properties of all nine galaxies in our sample. We discuss the results in more depth in the next section.

*Table 1. Basic Properties of the Galaxies*

| Source | z | $R_e$ | $v_c(R_e)$ | $\sigma_0$ | $v_c/\sigma_0$ | log $M_{baryon}$ | B/T | SFR | $f_{DM}(R_e)$ | $f_{gas}$ | sin i | Q | $t_{int}$ | FWHM Resolution |
|---|---|---|---|---|---|---|---|---|---|---|---|---|---|---|
| … | … | (kpc) | (km s$^{-1}$) | (km s$^{-1}$) | … | ($M_\odot$) | … | ($M_\odot$ yr$^{-1}$) | … | … | … | … | (hr) | (arcsec) |
| (1) | (2) | (3) | (4) | (5) | (6) | (7) | (8) | (9) | (10) | (11) | (12) | (13) | (14) | (15) |
| EGS_13035123 | 1.12 | 10.2 | 220 | 19 | 11.6 | 11.04 | 0.20 | 126 | 0.3 | 0.40 | 0.41 | 0.27 | 61 | 0.4–0.6 |
| zC_403741 | 1.45 | 2.6 | 206 | 60 | 3.4 | 10.59 | 0.70 | 60 | 0.1 | 0.38 | 0.47 | 1.00 | 12 | 0.4 |
| GS4_43501 | 1.61 | 4.9 | 259 | 60 | 4.3 | 10.92 | 0.05 | 53 | 0.3 | 0.45 | 0.88 | 0.58 | 22 | 0.5 |
| Q2343_BX610 | 2.21 | 4.0 | 295 | 60 | 4.9 | 10.94 | 0.12 | 140 | 0.3 | 0.62 | 0.63 | 0.47 | 58 | 0.2–0.5 |
| K20_ID7 | 2.23 | 7.9 | 322 | 60 | 5.4 | 11.27 | 0.04 | 101 | 0.6 | 0.64 | 0.85 | 0.37 | 33 | 0.4 |
| Q2346_BX482 | 2.26 | 5.8 | 293 | 67 | 4.4 | 10.91 | 0.02 | 80 | 0.6 | 0.63 | 0.87 | 0.52 | 18 | 0.4 |
| zC_405226 | 2.29 | 5.9 | 120 | 54 | 2.2 | 10.30 | 0.63 | 117 | 0.6 | 0.75 | 0.81 | 0.71 | 15 | 0.4 |
| D3a_15504 | 2.38 | 6.1 | 266 | 68 | 3.9 | 11.18 | 0.30 | 146 | 0.2 | 0.43 | 0.64 | 0.68 | 47 | 0.4 |
| D3a_6004 | 2.39 | 4.9 | 432 | 60 | 7.2 | 11.43 | 0.49 | 355 | 0.1 | 0.37 | 0.42 | 0.58 | 23 | 0.4 |

*Notes: Columns (1) to (13) are updated galaxy properties for our galaxies selected from the RC100 sample (Nestor et al. 2023): source name, redshift, effective radius ($R_e$), circular velocity at $R_e$ ($v_c(R_e)$), intrinsic velocity dispersion ($\sigma_0$), circular velocity to velocity dispersion ratio ($v_c/\sigma_0$), baryon mass, bulge-to-total ratio, star formation rate, dark matter fraction within $R_e$ ($f_{DM}(R_e)$), gas fraction ($f_{gas}$), sine of inclination angle (sin i), and Toomre Q parameter. All these values are consistent with RC100 (Nestor et al. 2023, Table B1) to within the 1σ uncertainty, except that for BX610 we updated SFR from far-infrared-based studies (e.g., Brisbin et al. 2019). Columns (14) and (15) list the on-source integration time and FWHM of the angular resolution of the IFU data used in this work.*

# 4. DISCUSSION AND INTERPRETATION

## 4.1. Observational Results for the Nine SFGs in this paper

In the following, we summarize our findings in all nine SFGs discussed in this paper. We have analyzed the remaining eight galaxies in the same manner as for Q2343_BX610 in Section 3.1. For most of these eight galaxies, at least two independent **DYSMAL** analyses were carried out, and for all, we carried out **kinemetry** analysis. The resulting maps and graphs for all eight galaxies are shown in the Appendix A (Figure A1 to Figure A17). Tables 2 and 3 list our results quantitatively.

1. In at least eight of our galaxies, we find evidence for significant velocity residuals with a $\pm$ signature along a specific direction. Assuming that these motions are in the plane of the galaxies, we deduce radial velocities between 30 ($\pm$14) and 120 ($\pm$47) km s$^{-1}$. In all but one of the galaxies (zC_405226) the motions deduced from **DYSMAL** forward modeling and **kinemetry are in excellent agreement.** This is shown in the last columns of Table 2 and the right-most panel of Figure 8.

2. In six SFGs this motion represents radial inflow **along or near the minor axis of the galaxy**. The relative angle between the major axis of the galaxy ($PA_{\text{major}}$) and the radial streaming ($PA_{\text{resid}}$) is near 90° (Table 3). As discussed in more detail in Section 3.1, in Q2343_BX610 we infer a $95 \pm 24$ km s$^{-1}$ inflow motion $\delta PA \sim -131°$ off $PA_{\text{major}}$, plausibly due to **radial streaming along a gas/stellar bar**. In Q2346_BX482 there is clear evidence for inflow along an axis $\delta PA \sim -150°$ (or +30°) from $PA_{\text{major}}$. There is no bar in this ring-like system, but two companions at distances of 1.9″ S and 3.3″ SE, and with an *H*-band flux ratio to the main galaxy of 0.3 and 0.23, respectively. The SE source has a velocity offset of +830 km s$^{-1}$ from the main galaxy, the S source is weak in Hα (velocity offset maybe $-200$ km s$^{-1}$).

It is unclear whether the perturbation by these lower mass companions could cause such a large disturbance in the velocity pattern. In GS4_43501 the residuals appear more complicated than a single gradient. If the gradient across the nucleus is real, the streaming could be either inwards or outwards. Finally, in zC_405226 there could be an inflow along the minor axis but the kinemetry results are quite uncertain.

3. The magnitude of the streaming motions increases with redshift (Figure 8, left panel). From $z \sim 1.3 - 2.3$ the average amplitude increases by about a factor of 2. The mean inflow velocity is $50 \pm 15$ km s$^{-1}$ for all nine SFGs. At $\langle z \rangle = 1.4$ we find $\langle v_r \rangle = 42 \pm 14$ km s$^{-1}$ and $\langle v_r/v_c \rangle = 0.19 \pm 0.06$, while at $\langle z \rangle = 2.3$ we find $\langle v_r \rangle = 91 \pm 33$ km s$^{-1}$ and $\langle v_r/v_c \rangle = 0.28 \pm 0.10$. Using the scaling relations of Tacconi et al. (2018, 2020), one would expect galaxies near $\log(M_{\text{baryon}}/M_\odot) = 10.8$ and on the MS to have an average molecular gas fraction of 0.3 at $z \sim 1.3$ and 0.54 at $z \sim 2.3$, an increase by a factor of 1.8. A further factor coming into play is that at higher $z$ and larger $f_{\text{gas}}$, $Q$ should drop, and $Q^{-\zeta}$ should increase the theoretically expected inflow velocity due to **Toomre disk instability and torques** (Section 4.2 below), close to the increase seen in Figure 8, and estimated in Table 2. Applying equation (11) with a~1.75, $\zeta=2$, and $\gamma=0$ (Section 4.2) we compute the theoretical expected values $v_r(Toomre)$ for the inward flow velocities. The middle panel of Figure 8 gives $v_r(obs)/v_r(Toomre)$ as a function of $v_r(obs)$.

4. All nine inferred values of $v_r$ in Table 2 (and certainly the six best ones) **are broadly consistent with the analytic predictions of inflow rates by global gravitational instabilities in gas-rich disks (see Section 4.1, Equation (11))**. The weighted mean ratio of $v_r(obs)/v_r(Toomre)$ in column 11 of Table 2 is 0.9, close to unity, for a~1.75, $\zeta=2$ and $\gamma=0$ in Equation (11). Recall our conclusion from the summary of the Q2343_BX610 section: if the flows are not axisymmetric but along a bar or spiral arms, the total inflow rates are likely smaller than $M_{\text{gas}}(R_e) \times v_r/R_e$.

*Table 2. Fitting Results for inflow velocities and kinemetry*

| Source | $z$ | $v_{resid}$ | $\delta v_{resid}$ | $v_{resid,SP}$ | $v_{resid,DL}$ | $v_r(obs) = v_{resid}/\sin i$ | $\delta v_r(obs)$ | $v_r(Toomre)$ | $\delta v_r(Toomre)$ | $\frac{v_r(obs)}{v_r(Toomre)}$ | $\delta \frac{v_r(obs)}{v_r(Toomre)}$ | $\frac{v_r}{v_c}$ | $\delta \frac{v_r}{v_c}$ | $\langle \frac{|A_1|}{B_1} \rangle$ | $\delta \langle \frac{|A_1|}{B_1} \rangle$ |
|---|---|---|---|---|---|---|---|---|---|---|---|---|---|---|---|
| | | (km s$^{-1}$) | (km s$^{-1}$) | (km s$^{-1}$) | (km s$^{-1}$) | (km s$^{-1}$) | (km s$^{-1}$) | (km s$^{-1}$) | (km s$^{-1}$) | | | | | | |
| (1) | (2) | (3) | (4) | (5) | (6) | (7) | (8) | (9) | (10) | (11) | (12) | (13) | (14) | (15) | (16) |
| EGS_13035123 | 1.12 | 17 | 7 | … | 17 | 42 | 14 | 31 | 8 | 1.37 | 0.57 | 0.19 | 0.06 | 0.04 | 0.012 |
| zC_403741 | 1.45 | 20 | 10 | … | … | 43 | 21 | 26 | 6 | 1.65 | 0.92 | 0.21 | 0.1 | 0.11 | 0.033 |
| GS4_43501 | 1.61 | 27 | 12 | … | 25 | 31 | 14 | 46 | 11 | 0.70 | 0.30 | 0.12 | 0.05 | 0.06 | 0.018 |
| Q2343_BX610 | 2.21 | 60 | 15 | 50 | 70 | 95 | 24 | 98 | 25 | 0.97 | 0.34 | 0.32 | 0.08 | 0.34 | 0.102 |
| K20_ID7 | 2.23 | 75 | 15 | 85 | 55 | 87 | 17 | 114 | 29 | 0.76 | 0.24 | 0.27 | 0.05 | 0.16 | 0.048 |
| Q2346_BX482 | 2.26 | 65 | 25 | 55 | 60 | 75 | 29 | 101 | 25 | 0.74 | 0.34 | 0.26 | 0.1 | 0.14 | 0.042 |
| zC_405226 | 2.29 | 82 | 30 | … | … | 101 | 37 | 59 | 15 | 1.73 | 0.77 | 0.84 | 0.31 | 1.27 | 0.635 |
| D3a_15504 | 2.38 | 50 | 25 | 60 | 55 | 78 | 39 | 43 | 11 | 1.82 | 1.02 | 0.29 | 0.15 | 0.23 | 0.069 |
| D3a_6004 | 2.39 | 50 | 20 | 80 | 55 | 118 | 47 | 51 | 13 | 2.31 | 1.09 | 0.27 | 0.11 | 0.17 | 0.051 |

*Notes: Columns (1) and (2) list galaxy name and redshift. Columns (3), (5) and (6) are the observed residual velocities revealed from the residual velocity maps after subtracting the circular motion kinematic models, measured by three of the authors (Section 2.1). Column (4) is the uncertainty of column (3). Column (7) is the inflow speed $v_r$ within galactic plane after accounting for the inclination. Column (8) is the uncertainty of $v_r$. Column (9) is the theoretical inflow velocity $v_r(Toomre)$ as described in Section 4.3 (Equation (11)), with its uncertainty in column (10). Column (11) and (13) are the ratios of $v_r/v_r(Toomre)$ and $v_r/v_c$, respectively, with their uncertainties in columns (12) and (14). Columns (15) and (16) list the kinemetry estimate of $\langle |A_1|/B_1 \rangle$ (probing $v_r/v_c$) and its uncertainty.*

*Table 3. Fitting Results for inflow vs. rotation directions*

| Source | z | $PA_{major,light}$ (deg) | $PA_{major,kin}$ (deg) | $PA_{resid}$ (deg) | $\delta PA$ (deg) | rotational motion on sky | $i$ (deg) | radial motions | shape of outer rotation curve | morphology |
|---|---|---|---|---|---|---|---|---|---|---|
| (1) | (2) | (3) | (4) | (5) | (6) | (7) | (8) | (9) | (10) | (11) |
| EGS_13035123 | 1.12 | -175 | -167 | -70 | 97 | Clockwise | 156 | Inflow along/near the minor axis | Peak-drop | Bulge & spiral |
| zC_403741 | 1.45 | -149 | -143 | -48 | 95 | Clockwise? | 152 | Inflow along/near the minor axis? | Peak-drop | Nucleus & ring |
| GS4_43501 | 1.61 | -32 | -32 | 75 | 107 | Clockwise | 118 | Inflow along/near the minor axis | Dropping | Nucleus & ring |
| Q2343_BX610 | 2.21 | -25 ± 8 | -25 ± 8 | -156 ± 8 | -131 | Counter-clockwise | 39 | Inflow along the bar at $PA\sim 25°$, 40° off minor axis | Peak | Bulge & disk |
| K20_ID7 | 2.23 | -164 | -156 | -54 | 102 | Clockwise | 122 | Inflow along/near the minor axis | Rising | Red nucleus & spiral |
| Q2346_BX482 | 2.26 | 121 | 111 | -39 | -150 | Counter-clockwise | 60 | Inflow along the axis different from the minor axis | Rising | Ring, companions†. |
| zC_405226 | 2.29 | 116 | 150 | 240 | 90 | Clockwise | 126 | Inflow along/near the minor axis | Flat | Clumpy spiral with bulge |
| D3a_15504 | 2.38 | 164 | 155 | 67 | -88 | Counter-clockwise? | 40 | Inflow along/near the minor axis, but if clockwise outflow | Dropping | Disk, companion‡. |
| D3a_6004 | 2.39 | 37 | 167 | 52 | -115 | Counter-clockwise | 25 | Inflow near the minor axis | Peak | Big bulge & ring, companion‡‡. |

*Notes: Column (3) lists the PA of the major axis of stellar light ($PA_{major,light}$, positive east of north, the same for other PAs). Column (4) lists the PA of the line of minimum and maximum velocities ($PA_{major,kin}$, indicating the kinematic major axis). Column (5) lists the PA of the line of maximum and minimum velocities in the velocity residual map ($PA_{resid}$, indicating noncircular motion), followed by the angle difference $\delta PA = (PA_{resid} - PA_{major,kin})$ in Column (6). Descriptions of galactic rotational motion, inclination, radial motions, classification of outer rotation curves and morphology are given in columns (7) to (11), respectively. †Q2346_BX482 has two companions: companion a is ~1.9" away at SW and companion b is ~3.3" away at SE, see Section 4.2. ‡D3a_15504 has an interacting companion ~1.5" away at NW, with a mass ratio of about 30:1, and velocity offset about −50 km s$^{-1}$. ‡‡D3a_6004 has a companion ~1.7" away at SE, with a mass ratio of about 50:1.*

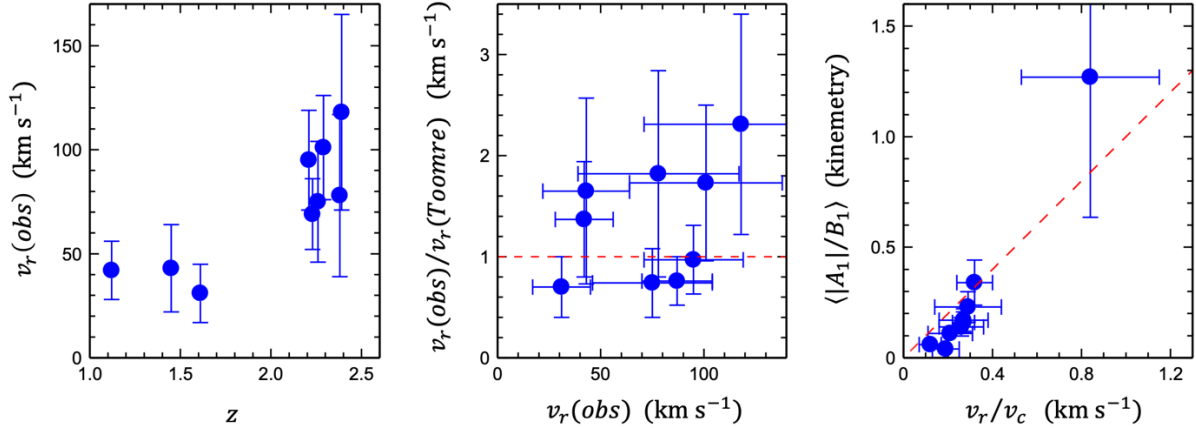

Figure 8. Correlation between derived properties of the nine galaxies. For the computation of the theoretically expected converging flow speed (Table 2) we use equation 11 with a~1.75, ζ=2 and γ=0. We also use for each galaxy the estimates of gas fraction and Q as given in Table 1.

### 4.2. Caveats: Neighbors, Tidal Effects and Warps

We identified small satellite galaxies at a projected distance of < 30 kpc for 3 of our 9 galaxies (Q2346_BX482, D3a_15504, and D3a_6004). From the deep ALMA observations, we additionally identified two potential satellites/neighbors for Q2343_BX610 at a larger distance of >100 kpc (13″ and 20″). Depending on the spectral energy distribution models we adopt, the mass ratios are ~1:3–1:8 and ~1:2–1:1 with respect to Q2343_BX610's mass (M. Lee et al. 2023, in preparation). Following the exercise in Genzel et al. (2017, 2020), we introduce a rough estimator of the impact of the satellites on the central galaxy. Equation (8.14) in Binney & Tremaine 2008 (Chapter 8) describes the Jacobi (or Hill) radius ($R_J$) for a singular isothermal sphere ($M \sim R$). For a test particle with mass $M_2$ at a distance $R$ with the interior mass of the central galaxy ($M_1(R)$), the Jacobi radius is defined by

$$R_J = R \times \left(\frac{M_2}{3M_1(R)}\right)^{1/3} = 3 \text{ kpc} \times \left(\frac{M_2/M_1}{0.05}\right)^{1/3} \times \left(\frac{R_{12}}{12 \text{ kpc}}\right)$$
(5).

We take the ratio between this Jacobi radius ($R_J$) at effective radius $R_e$ (where $M_1(R_e) = 0.5 M_1$) and the distance between the central galaxy (1) and the neighboring galaxy (2), ($R_J/R_{12}$). This ratio ranges between 0.02 and 0.2 (median ~0.1) for all our galaxies. These values are small enough and support the conclusions of Genzel et al. (2017, 2020) that at least currently identified neighbors/satellites do not play a significant role as strong kinematic perturbers.

In addition to interactions, **warps** can be important in outer galactic disks and are frequently observed in the outer HI layers of z~0 galaxies (including the Milky Way, Levine, Blitz & Heiles 2006; van der Kruit & Freeman 2011). Such warps can generate apparent large line-of-sight velocity changes that would be interpreted as noncircular motions in a planar disk model (e.g., Peterson et al. 1978). Theoretically, this type of **buckling or firehose instability** (with a predominant $m = 2$ mode) can occur in galaxy disks with surface density $\Sigma$, with radial wavelength $\lambda$ less than

$$\lambda \leq \lambda_J = \frac{\sigma_x^2}{G\Sigma}$$
(6).

Here $\sigma_x$ is the in-plane velocity dispersion (Binney & Tremaine 2008, Chapter 6.6.1; Toomre 1964). Toomre (1964) and Merritt & Sellwood (1994) have shown that the system must be sufficiently cold in the vertical direction for the instability to grow. This means that the z-scale height $h_z$ has to be smaller than

$$h_z < \frac{\sigma_z^2}{G\Sigma} \text{ such that } \frac{\sigma_z}{\sigma_x} < a_{crit} \sim 0.3....0.6 \quad (7).$$

The current data for high-z galaxies suggest that the velocity dispersion **ellipsoid is isotropic** (Genzel et al. 2011; van Dokkum et al. 2015; Wisnioski et al. 2015; Übler et al. 2019), $\sigma_x = \sigma_z$, such that warping should be suppressed.

If the warp has a sufficiently high amplitude, it could indeed introduce a radial dependence of the peak rotation velocity along the major axis. If the dominant mode is uneven ($m = 1$ or $m = 3$, as in the Milky Way, Levine, Blitz & Heiles 2006), warps would also introduce the same sign of the change in the absolute value of the peak rotation curve on the blue- and the redshifted side of the galaxy. This could mimic a radial decrease (or increase) in the rotation curve, with equal probability. If the mode is even ($m = 2$), however, one should observe rotation curves that increase on one side, and decrease on the other. We do not observe such rotation curves in any of the galaxies presented here, nor in the underlying larger samples RC100 (Genzel et al. 2020; Nestor et al. 2023). Again, this speaks against high amplitude warps as the cause of our second-order velocity residuals.

### 4.3. Summary of Theoretical Background

#### 4.3.1. Analytical Estimates Based on the Toomre Disk Instability

**Star-forming galaxies at the peak of the star-forming activity are rich in cold molecular gas** ($f_{gas} = M_{gas}/(M_{gas} + M_*) \sim 0.3 \cdots 0.75$, see Tacconi et al. 2020 and references therein). All nine SFGs discussed in this paper are comparably gas-rich. As a result, in contrast to the situation at $z \sim 0$ (e.g., Leroy et al. 2013), high-$z$ SFGs are **globally gravitationally unstable** (Lin & Pringle 1987; Noguchi 1999; Immeli et al. 2004; Förster Schreiber et al. 2006; Bournaud et al. 2007; Elmegreen et al. 2007; Escala & Larson 2008; Genzel et al. 2008; Bournaud & Elmegreen 2009; Dekel, Sari & Ceverino 2009; Förster Schreiber et al. 2009; Krumholz & Burkert 2010). Their Toomre parameter $Q$ is $\leq 1$ (see Column 13 of Table 1). All spatial scales between the classical Jeans length $\lambda_J = c_s^2 / G\Sigma \sim O(1\,pc)$ for speed of sound $c_s$ and surface density $\Sigma$ (gas and stars), and the Toomre length

$$\lambda_T = Q \times \pi^2 G\Sigma \times \left(v_{rot}/R_{disk}\right)^{-2} \sim f_{gas} R_{disk} \sim 2\,\text{kpc} \quad (8)$$

become gravitationally unstable, resulting in the formation of **large, massive clumps** of mass $M_T \sim f_{gas}^2 M_{disk} \sim 10^8 \cdots 10^{9.5}$ M$_\odot$. Larger transitory structures, such as **spiral features, or bars** can form as well. Our estimates of $Q < 1$ in the observed disks (see Table 2) are indicative of Toomre instability, but with the caveat that the observed disks are in the nonlinear regime of gravitational instability ($Q<1$) while analytical Toomre theory strictly refers to linear instability ($Q \sim 1$).

Large-scale disk instabilities or other perturbations in the disk induce gravitational torques that cause the transport of angular momentum outward, which is compensated by mass transport inward. As a result, giant gas/star-forming clumps migrate toward the disk center, possibly forming early, star-forming bulges (Noguchi 1999; Immeli et al. 2004; Bournaud et al. 2007; Elmegreen, Bournaud & Elmegreen 2008; Ceverino, Dekel & Bournaud 2010; Dekel et al. 2022). In addition, the disks are fed from the CGM with fresh gas in non-axisymmetric streams (Dekel et al. 2009), which also may introduce torques and strong inward motions, perhaps also resulting in bar formation (Danovich et al. 2015). The inter-clump torques increase the velocity dispersion and scale height of the disk. The associated radial velocities have been estimated analytically in several different ways. These include energy-conserving radial transport in a viscous disk, clump migration due to clump encounters, dynamical friction, and radial transport of a ring torqued by tightly wound spiral arms (e.g., Dekel, Sari & Ceverino 2009; Krumholz & Burkert 2010; Dekel & Krumholz 2013; Dekel et al. 2013; Rathaus & Sternberg 2016; Krumholz et al. 2018; Dekel et al. 2020). Broadly, the transport timescale (e.g., the viscous (vis), or dynamical friction (df)) time can be written as

$$t_{vis} \sim t_{df} \sim \left(\frac{v_c}{\sigma_0}\right)^2 \times t_{dyn}(R) \quad (9),$$

where $t_{dyn} = R/v_c(R)$ is the dynamical time at $R$, $\sim 15\,Myr$ for the SFGs in Table 1. The $z$-scale height of the disk becomes comparable to the Toomre scale, such that

$$\left(\frac{h_z}{R_{disk}}\right) = \left(\frac{\sigma_0}{v_c}\right) = \frac{Q \times f_{gas}}{\sqrt{2\ldots 3}} \quad (10).$$

At $z \sim 2$ typical velocity dispersions are 40–70 km s$^{-1}$ (Übler et al. 2019), and $v_c/\sigma_0 \sim 4$.

The resulting inflow velocity scales as

$$v_r = \frac{R_{disk}}{t_r} \sim bv_c \times \left(\frac{\sigma_0}{v_c}\right)^2 \quad (11).$$

Depending on assumptions $b = a \times Q^{-\zeta} \times \left(\frac{\sigma_0}{v_c}\right)^\gamma_{0.25}$ where $a \sim 1\text{-}4$, $\zeta = 1\text{-}4$, and $\gamma = 0\text{-}1$ (Dekel, Sari & Ceverino 2009; Krumholz & Burkert 2010; Dekel et al. 2013, 2020). The cases $a=4$ and $\zeta=4$ refer to the radial velocity of individual clumps, NOT to the azimuthally averaged radial flow. As such they do not pertain to the case explored in this paper. Hence, we take an average between Krumholz & Burkert (2010) and Krumholz et al. (2018), which indeed refer to azimuthally averaged (axisymmetric) flow: **$a \sim 1.75$, $\zeta = 2$, and $\gamma = 0$**. We note that the **variation of these parameters in the different theory papers leads to very substantial differences in expected output parameters by a factor of 2 or more**.

The above discussion indicates that in the disk instability regime, one should expect large changes in the magnitude of $v_r/v_c$ with $z$. Taking the scaling relations between $\sigma_0$ and $z$ by Übler et al. (2019), and between $f_{gas}$ and $z$ by Tacconi et al. (2018, 2020), $Q$ should decrease from $\sim 1.05$ to $\sim 0.45$, $f_{gas}$ from 0.17 to 0.62, and $v_r/v_c$ from 0.02 to 0.25–0.3, between $z = 0.6$ and 2.6. For $\zeta=2$, $v_r/v_c \sim (f_{gas})^2$.

### *4.3.2. Bar and Spiral Arm Flows*

Non-axisymmetric perturbations in the gravitational potential effectively induce radial gas flows as well. For instance, Roberts, Huntley and van Albada (1979) report on steady-state gas-dynamical studies, previously limited to tightly wound normal spirals, to include barred and open-armed normal spirals (see also van Albada & Roberts 1981, Athanassoula 1992, Bournaud & Combes 2002, Rodriguez-Fernandez & Combes 2008). Roberts et al. find that "…the steady-state response of the gas (non-self-gravitating) to a 5-10% perturbing potential that is bar-like in the inner parts and spiral-like in the outer parts is found to be strong and capable of inducing the formation of large-scale gaseous density waves and shocks in the bar and along the spiral arms. **Highly oval streamlines characterize the gas circulation in the inner regions of the disk where large noncircular motions** are of the order of $50 \text{ km s}^{-1}$ to $150 \text{ km s}^{-1}$ (see also Peterson et al. 1978). Strong velocity gradients in the gas flow are particularly pronounced across the bar near the shock…".

### *4.3.3. Simulation Results*

Slyz et al. (2002) used 2D **hydrodynamical simulations of viscous disk evolution**, including star formation, to test the analytical Lin & Pringle (1987) model that exponential stellar disks arise naturally, if star formation proceeds on the same timescale as the viscous redistribution of mass and angular momentum. They showed that this conjecture is indeed true, regardless of the disk's initial gas distribution and rotation curve. They propose that there exists a strong physical link between star formation and viscosity and that the viscous timescale is the natural timescale for star formation. More recently Forbes, Krumholz & Burkert (2014) used GIDGET, an axisymmetric disk evolution code, especially designed to study the viscous evolution of gravitationally unstable disks and demonstrated that there is an intimate balance between external accretion, gravitational instability, star formation, and radial gas flow through the disk. Goldbaum, Krumholz and Forbes (2015) performed a set of high-resolution, fully 3D simulations of isolated disks, including star formation but neglecting stellar feedback. The gas disks become gravitationally unstable with substantial turbulent velocity dispersions that lead to radial inflow. Indeed, the mass transport inward is equal to the star formation rate, which for their Milky Way-like galaxies is of order $1 \, M_\odot \text{ yr}^{-1}$. In a subsequent paper (Goldbaum, Krumholz & Forbes 2016) they explored the role of stellar feedback, which has a strong effect on the multiphase structure of the interstellar medium and reduces star formation rates by a factor of 5. For their low-, fiducial-, and high gas fraction simulations they find star formation rates of 0.3, 2, and $10 \, M_\odot \text{ yr}^{-1}$. The average radial mass transport inward is 0.1 and $1 \, M_\odot \text{ yr}^{-1}$ for the low- and fiducial gas fractions, roughly a factor 2 smaller than the star formation rates. Unfortunately, their high gas fraction simulations did not run long enough to reach a stable inflow rate.

Early **cosmological simulations** of the formation and evolution of gas-rich disks showed in-spiral of massive clumps through **dynamical friction**, following the triggering of disk instability (e.g., Noguchi 1999; Immeli et al. 2004; Bournaud et al. 2007; Ceverino, Dekel & Bournaud 2010). The authors find of order 10%-20% of the disk gas falls into the center by clump migration on timescales of 10 dynamical times, $t_d = 50$ Myr. Naturally, these simulations had to make simplified feedback prescriptions and had limited spatial resolution, which resulted in artificially large and long-lived clumps.

More recent simulations with better resolution and updated feedback recipes, and appropriately high gas fractions are required in order to test the conjecture that the gas infall rates are on average similar to the star formation rates. Adopting this assumption the gas infall rate is given by

$$\frac{dM}{dt} = \beta \frac{M_{gas} \times v_r}{R_{disk}} = SFR = \frac{M_{gas}}{t_{depl}} \qquad (12),$$

where SFR is the star formation rate, $t_\text{depl}$ is the gas depletion timescale and $\beta$ denotes the fraction of the disk area affected by the converging flow. The inflow velocity is then

$$v_r = \frac{R_{disk}}{\beta \times t_{depl}} = 40 \text{ km/s} \times \left(\frac{R_{disk}}{4 \text{ kpc}}\right) \times \left(\frac{5 \times 10^8 \text{ yr}}{t_{depl}}\right) \times \left(\frac{0.2}{\beta}\right)$$

(13).

As already stated above, in this estimate the high inflow rates and large inflow velocities for high-z galaxies are coupled to the large star formation rates, which in turn result from large gas masses and small depletion timescales, compared to present-day galaxies.

The VELA zoom-in cosmological simulations (D. Dutta Chowdhury et al. 2023, in preparation) do show significant non-axisymmetric radial velocities, with large angular variations, while the angular-averaged radial velocities are lower than predicted by the analytic models. On the other hand, in recent AREPO simulations (zoom-in resimulations of TNG100; Pillepich et al. 2018, 2019), S. Pastras et al. (2023, in preparation) present cosmological simulations with improved resolutions of ~2.5 pc and gas fractions of ~45% at $z \sim 3$. They find large non-axisymmetric radial velocities (up to ~$60 \text{ km s}^{-1}$) and large-scale spiral arm formation, which transport gas efficiently inward and form nuclear concentrations. The Pastras et al. work is among the highest-resolution cosmological simulations achieved so far, and may indicate the importance of resolving small-scale structures.

Given these uncertainties in the state-of-the-art simulations, a decisive theoretical interpretation will have to await future progress.

## 5. SUMMARY

We have presented high-quality and high-resolution (0.2″–0.5″) imaging spectroscopy of Hα, [NII] 6583 and CO 3–2/4–3 in nine moderately large, modest, or low-inclination rotating disks near the peak of cosmic galaxy formation (z~1-2.5). Our parent sample is RC100, which represents a fair sampling of massive ($\log M_*/M_\odot > 9.5-11.2$) SFGs near the main-sequence of star formation (Genzel et al. 2020; Price et al. 2021; Nestor et al. 2023).

From RC100 we selected disks with no or little perturbation by nearby (massive) companions. All disks are rotating and we have used the high-quality rotation curves, along with priors from stellar and gas measurements and HST imagery, to carry out forward modeling to construct model velocity and velocity dispersion maps for pure rotating systems. We use the modeling tool **DYSMAL** to include the effects of beam smearing and instrumental resolution in order to calculate 3D data cubes. Subtracting the model velocity and velocity dispersion maps from those observed allows studying **second-order, velocity residuals due to radial streaming motions** in our program galaxies. Independently we use **kinemetry** in the velocity maps directly to infer **model-independent** evidence for deviations from pure circular motions.

**All nine galaxies exhibit significant, or large non-tangential motions, assuming that the gas motions are in the disk planes**. For six of the nine galaxies, we clearly detect large converging flows (inflows) (typically 75 km s$^{-1}$, or 0.26 of the rotational component) along the minor axes of the galaxies with both analysis techniques. Such a pattern would be expected for axisymmetric radial flows if velocity dispersions are large, gas fractions are high, and if the galaxies are very Toomre unstable (Noguchi 1999; Immeli et al. 2004; Bournaud et al. 2007). All these criteria are fulfilled by the nine galaxies (Columns 5, 11 and 13 of Table 1). In two galaxies we detect converging flows but along an axis offset from the minor axis (Q2343_BX610, Q2346_BX482). In Q2343_BX610 we find that this axis matches a gaseous-stellar bar. In GS4_43501 the signature of a converging flow is only convincing close to the center.

We cannot exclude that the measured deviations from tangential motion are caused (in part) by **substantial non-planar motions**. However, the large velocity dispersions and large cold gas columns in all of our nine galaxies make the turbulent, high column density, thick disks less prone to buckling instabilities (Toomre 1964; Merritt & Sellwood 1994), which would cause warps. The fact that we see no significant examples of outer rotation curves with asymmetric shapes relative to the center in RC100 (dropping on one side, and rising on the other, see Genzel et al. 2020; Nestor et al. 2023) also speaks against $m = 2$ induced warps.

**All SFGs in our sample are rich in dense interstellar molecular gas** ($f_{\rm gas} = M_{\rm gas}/(M_{\rm gas} + M_*) \sim 0.35-0.75$, Tacconi, Genzel & Sternberg 2020). As a result, theory predicts the gas disks to be locally and globally unstable to gravitational collapse from the Jeans length to the Toomre scale, which is $\sim 0.3 - 0.7 \times R_{\rm e} \sim 2$ kpc. As a result large clumps form and collapse to star clusters, which have been studied in many high-resolution HST images of the last two decades. These clumps and other large-scale structures (such as external converging streams from the CGM) plausibly exert large torques in the gas disk, resulting in angular momentum transport.

Our observations imply that **gas at $R_{\rm e}$ is transported rapidly to the central regions on ~5–15 dynamical time scales**, between 40 and 200 million years. **The total amount of gas affected by this inflow is much smaller than the total gas in the disk if the flow happens only in a small sector of the disk, such as along a bar, or a spiral arm.** It approaches the gas mass only if the flow is axisymmetric around the entire circumference of the disk. Independent evidence for significant gas transport comes from the gas/dust concentrations or large central extinctions in K20_ID7, Q2343_BX610, and GS4_43501. We expect that this transport will trigger active nuclear star formation, rapid bulge growth, and efficient feeding of central supermassive black holes. Here we note that submillimeter/millimeter or mid-IR high-resolution data (from NOEMA, ALMA, and JWST) will be essential, since the rest-frame optical data (from Hα and stellar optical light) are strongly affected by extinction.

## ACKNOWLEDGEMENTS

The authors thank the anonymous referee for their helpful comments. This research is based in part on observations of an IRAM Legacy Program carried out with NOEMA, operated by the Institute for Radio Astronomy in the Millimetre Range (IRAM), which is funded by a partnership of INSU/CNRS (France), MPG (Germany), and IGN (Spain), and in part on observations collected at the European Southern Observatory under ESO programs 092A-0091, 093.A-0079, 093.A-0079, 094.A-0217, 095.A-0047, 096.A-0025, 097.A-0028, 098.A-0045, 099.A-0013, 0100.A-0039, and 0101.A-0022. We make use of ALMA data: ADS/JAO.ALMA#2013.1.00059.S, ADS/JAO.ALMA#2017.1.01045.S and ADS/JAO.ALMA#2019.1.01362.S. ALMA is a partnership of ESO (representing its member states), NSF (USA) and NINS (Japan), together with NRC (Canada), MOST and ASIAA (Taiwan), and KASI (Republic of Korea), in cooperation with the Republic of Chile. The Joint ALMA Observatory is operated by ESO, AUI/NRAO and NAOJ.

This research was supported by the Excellence Cluster ORIGINS, which is funded by the Deutsche Forschungsgemeinschaft (DFG, German Research Foundation) under Germany's Excellence Strategy – EXC – 2094 – 390783311. R.H.-C. thanks the Max Planck Society for support under the Partner Group project "The Baryon Cycle in Galaxies" between the Max Planck Institute for Extraterrestrial Physics and the Universidad de Concepción. R.H-C also acknowledges financial support from Millenium Nucleus NCN19058 (TITANs) and support by the ANID BASAL projects ACE210002 and FB210003.

# APPENDIX A: DATA AND RESULTS FOR EIGHT ADDITIONAL SFGS

## A.1 EGS_13035123 (z=1.1)

Figure A1 shows the HST $V$, $I$, $H$ bands, CO, [CI], stellar mass, attenuation ($A_V$) and dust continuum maps of EGS_13035123. We obtained the CO, [CI] and dust continuum maps with NOEMA.

Figure A2 shows the 1D rotation velocity and dispersion profiles, 2$^D$ velocity and dispersion maps, and their residual maps.

Figure A3 illustrates the kinemetry diagnostic plot similar to that in Figure 7. Here the spiral arms are used to determine the direction of the rotational motion on the sky.

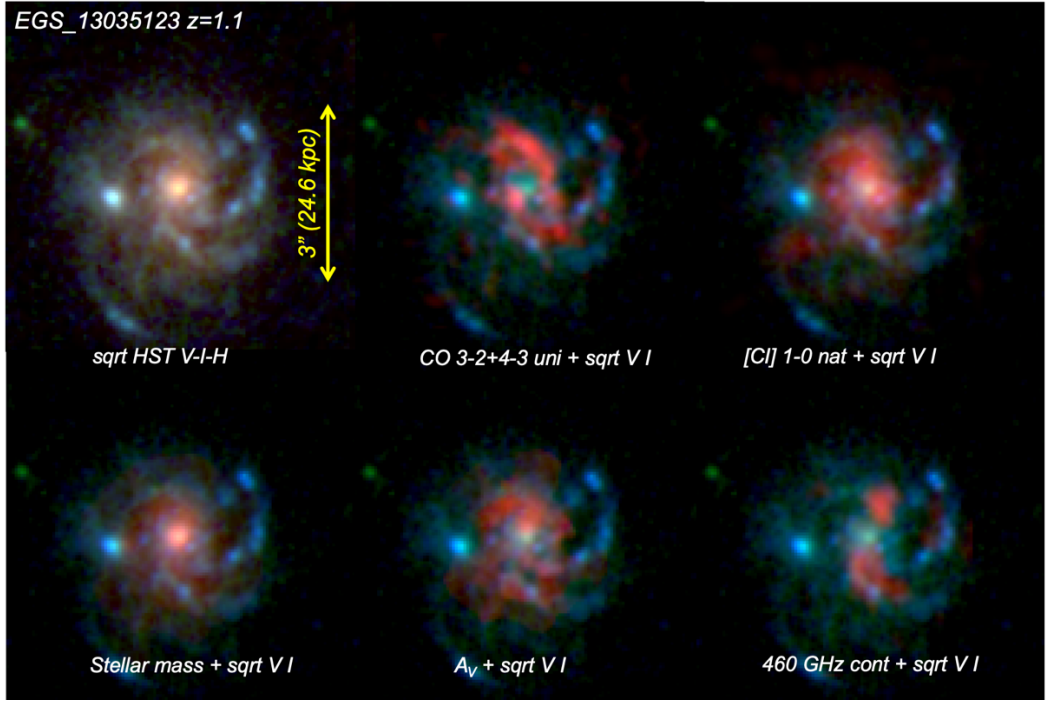

*Figure A1: Images of continuum and integrated line emission in EGS_13035123 (z = 1.1).*

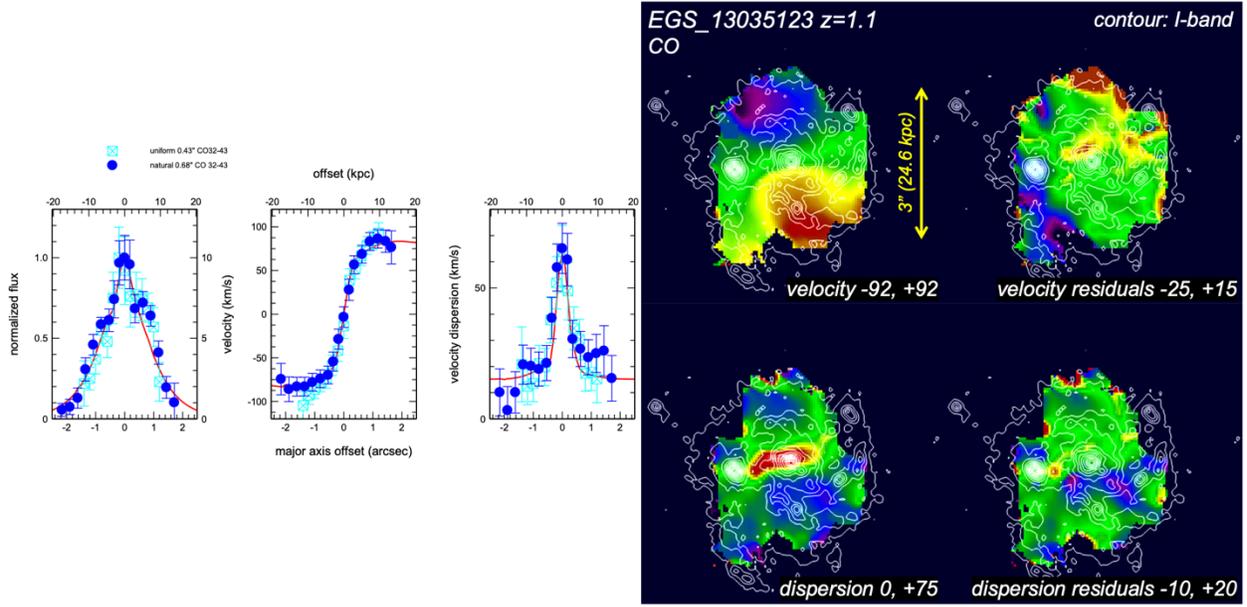

*Figure A2: Left: CO 3–2/4–3 intensity, velocity, and velocity dispersion 1D profiles and kinematic fitting results along $PA_{major}$ (uniform weighting data: cyan crossed squares, FWHM 0.43"; natural weighting data: filled blue circles, FWHM 0.68"). Right: Velocity, velocity dispersion (top) and residual maps (bottom) of CO 4–3/3–2 in EGS_13035123.*

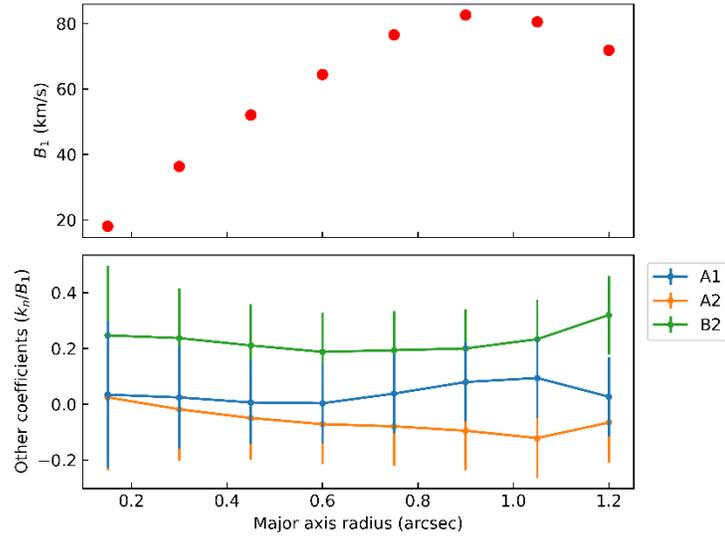

*Figure A3: Kinemetry of CO velocity maps in EGS_13035123.*

### *A.2 zC_403741 (z=1.45)*

Figures A4 and A5 show the 1D and 2D kinematic residual maps and kinemetry diagnostic plots for zC_403741. The direction of the rotation is not clear but likely clockwise (as stated on Table 3 with a question mark) based on extinction.

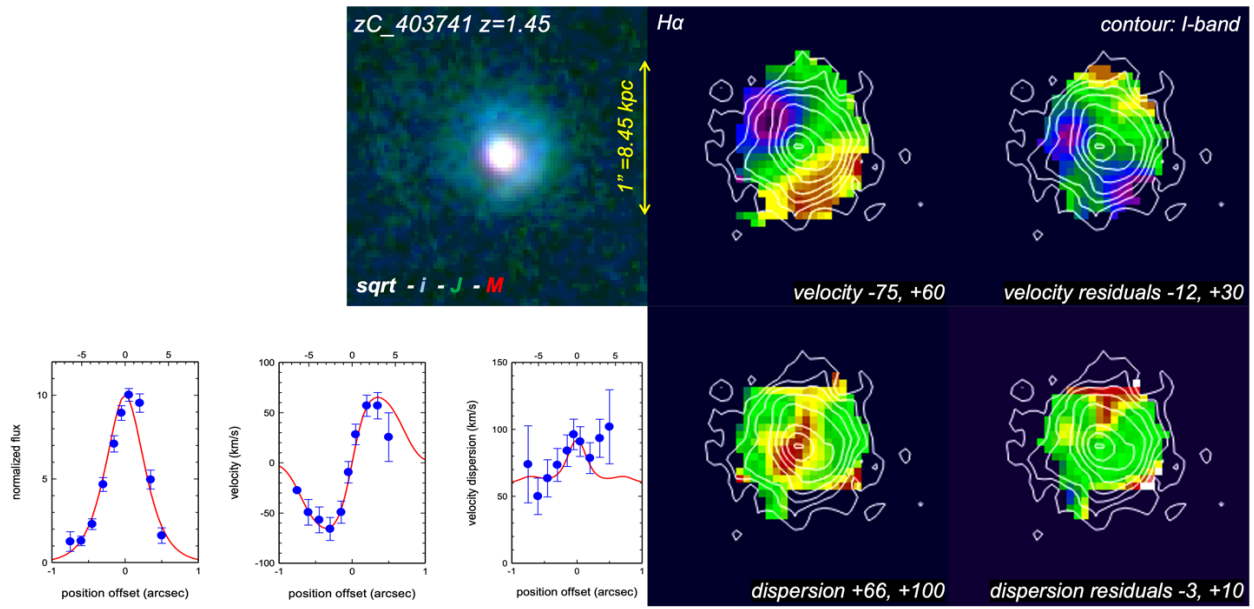

*Figure A4: Top and Right: Images of continuum and Hα velocity, velocity dispersion, and their residuals in zC_403741 (z=1.45). Bottom Left: 1D Hα intensity, velocity, and velocity dispersion cuts along $PA_{major}$, as well as the best-fit **DYSMAL** model (red line).*

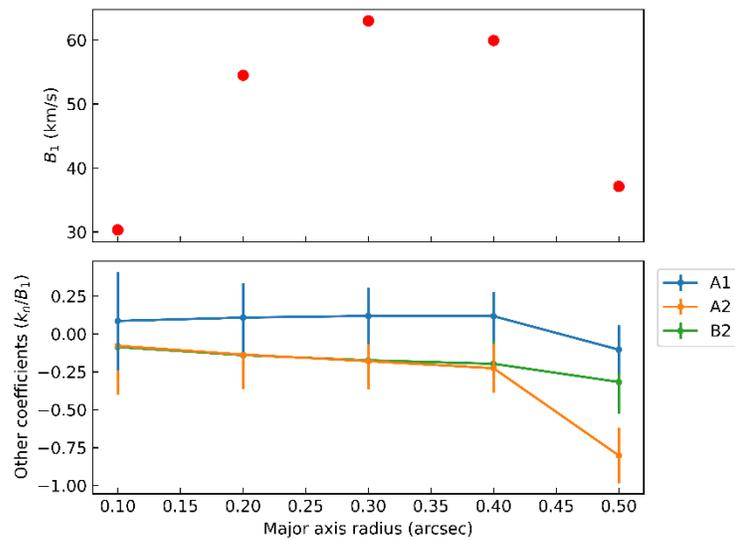

*Figure A5: Kinemetry of Hα velocity maps in zC_403741.*

### *A.3 GS4_43501 (z=1.61)*

Figures A6 and A7 show the 1D and 2D kinematic residual maps and kinemetry diagnostic plots for GS4_43501. Extinction is used to determine which side is closer to us and hence the direction of the rotational motion on the sky.

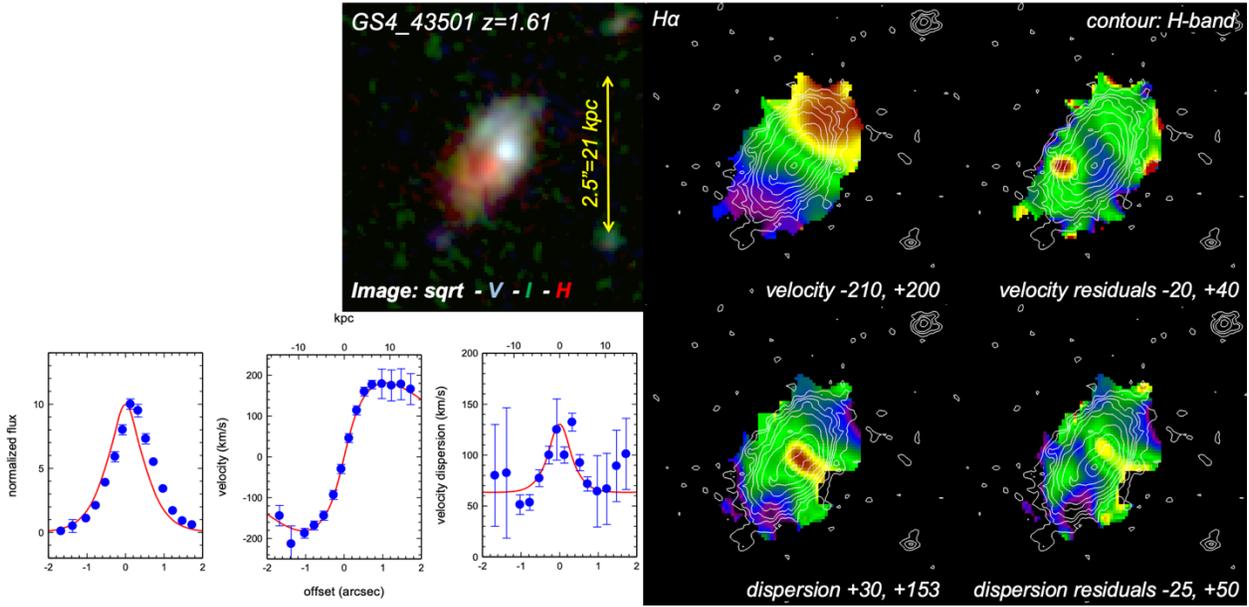

*Figure A6: Top and Right: Images of continuum and Hα velocity, velocity dispersion, and their residuals in GS4_43501 (z=1.61). Bottom Left: 1D Hα intensity, velocity and velocity dispersion cuts along $PA_{major}$, as well as the best-fit **DYSMAL** model (red line).*

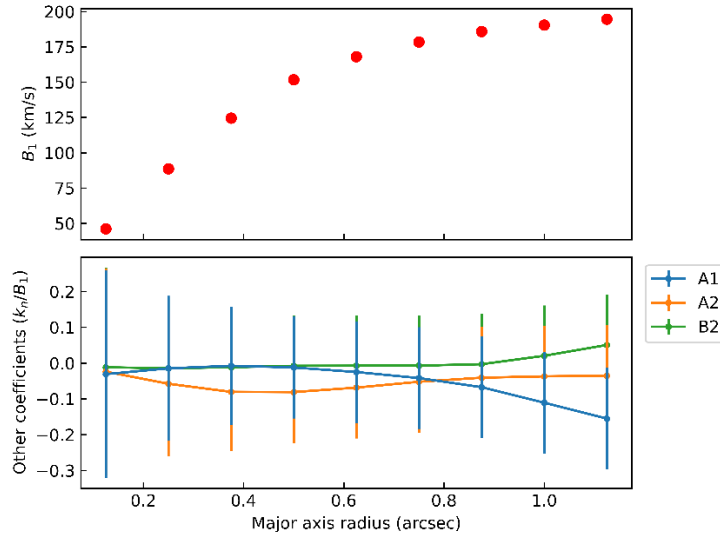

*Figure A7: Kinemetry of Hα velocity maps in GS4_43501.*

## *A.4 K20_ID7 (z=2.23)*

Figures A8 and A9 show the 1D and 2D kinematic residual maps and kinemetry diagnostic plots for K20_ID7. The spiral arms are used to determine the direction of the rotational motion on the sky.

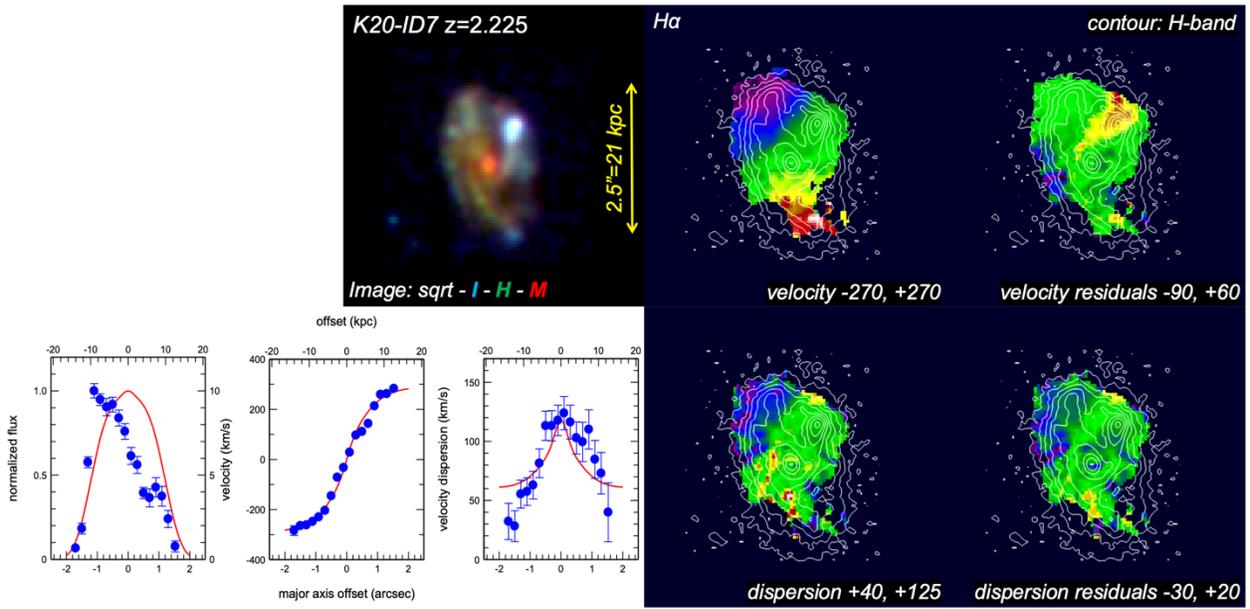

*Figure A8: Top and Right: Images of continuum and Hα velocity, velocity dispersion, and their residuals in K20_ID7 (z=2.23). Bottom Left: 1D Hα intensity, velocity and velocity dispersion cuts along $PA_{major}$, as well as the best-fit **DYSMAL** model (red line).*

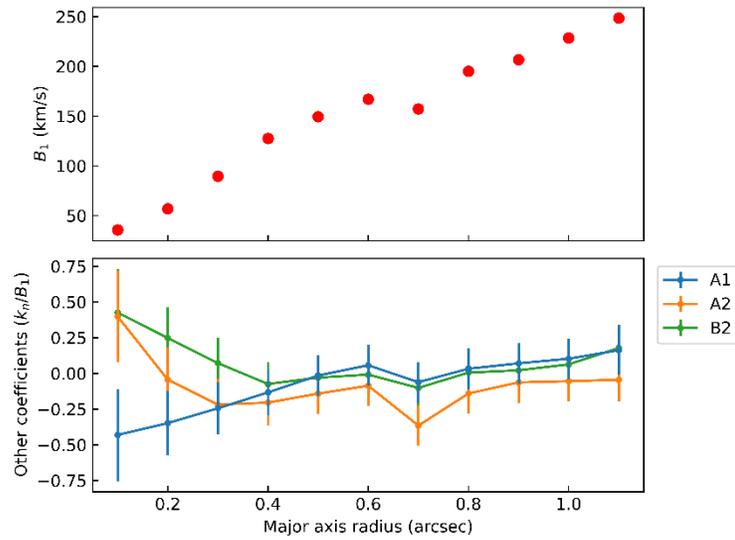

*Figure A9: Kinemetry of Hα velocity maps in K20_ID7.*

### *A.5 Q2346_BX482 (z=2.26)*

Figures A10 and A11 show the 1D and 2D kinematic residual maps and kinemetry diagnostic plots for Q2346_BX482. Extinction is used to determine the orientation of the galaxy.

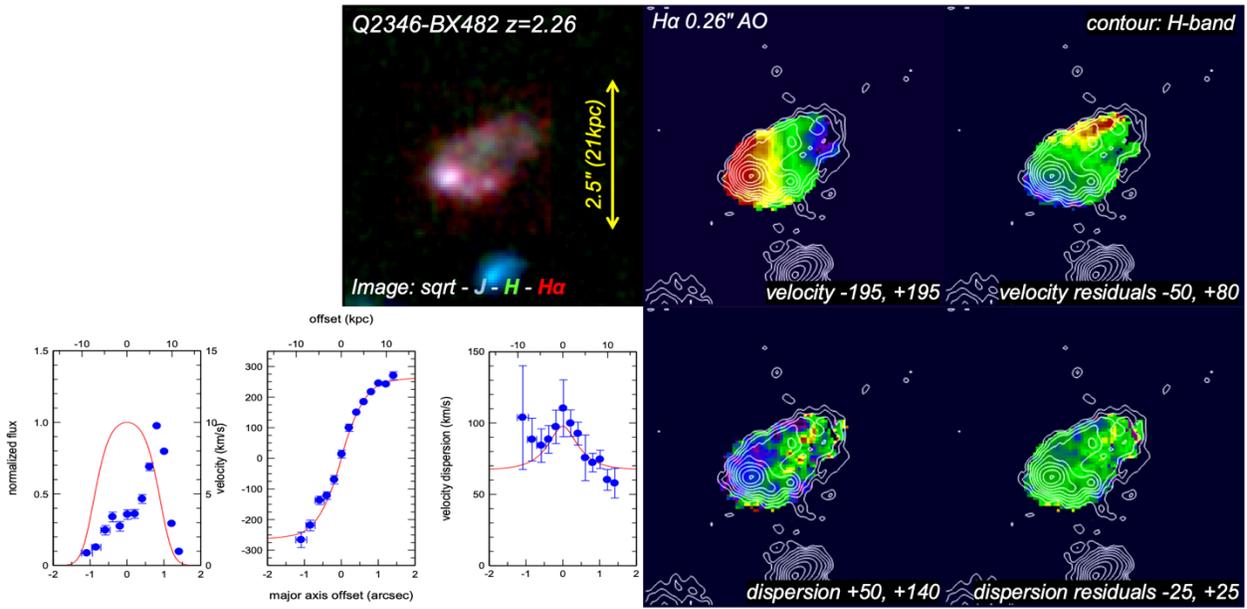

*Figure A10: Top and Right: Images of continuum and Hα velocity, velocity dispersion, and their residuals in Q2346_BX482 (z=2.26). Bottom Left: 1D Hα intensity, velocity and velocity dispersion cuts along $PA_{major}$, as well as the best-fit **DYSMAL** model (red line).*

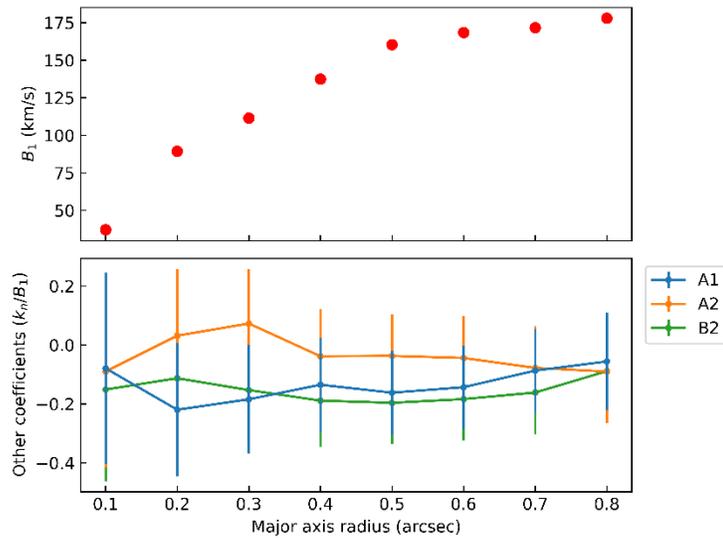

*Figure A11: Kinemetry of Hα velocity maps in Q2346_BX482.*

### *A.6 zC_405226 (z=2.29)*

Figures A12 and A13 show the 1D and 2D kinematic residual maps and kinemetry diagnostic plots for zC_405226. Extinction is used to determine the orientation.

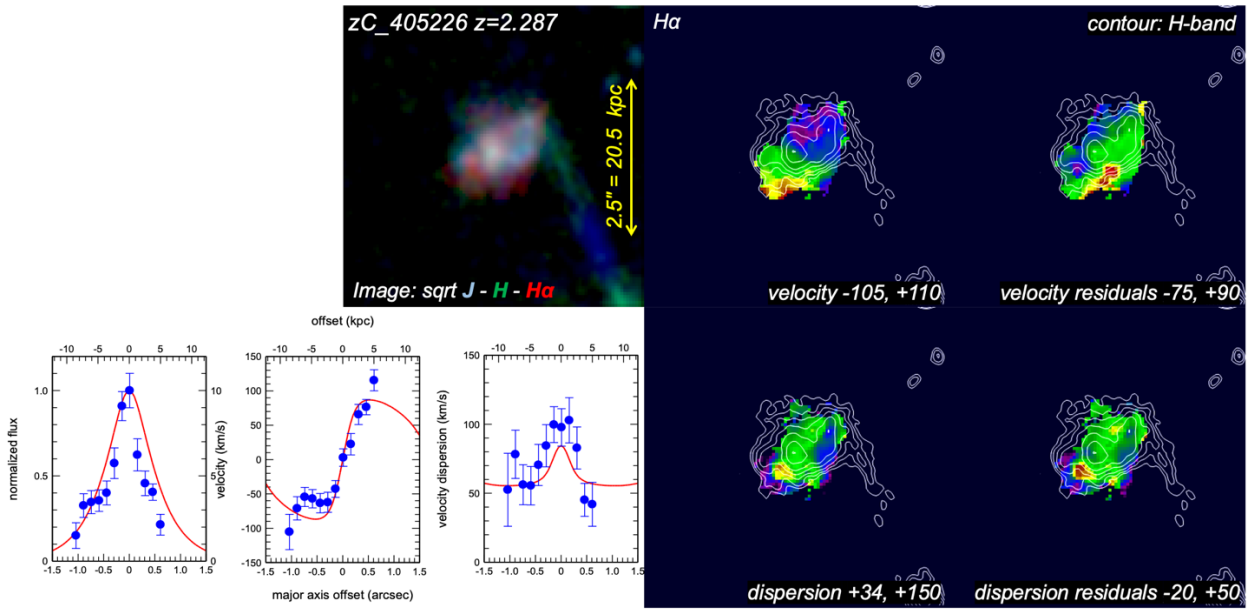

*Figure A12: Top and Right: Images of continuum and Hα velocity, velocity dispersion, and their residuals in zC_405226 (z=2.29). Bottom Left: 1D Hα intensity, velocity and velocity dispersion cuts along $PA_{major}$, as well as the best-fit **DYSMAL** model (red line).*

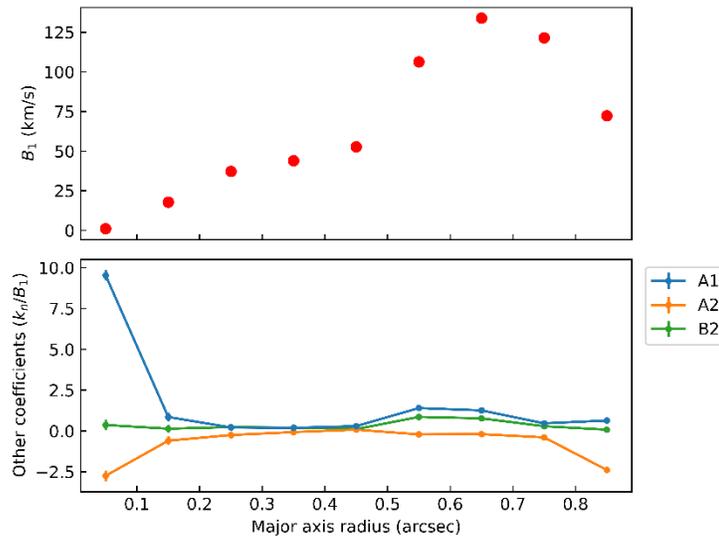

*Figure A13: Kinemetry of Hα velocity maps in zC_405226.*

## *A.7 D3a_15504 (z=2.38)*

Figures A14 and A15 show the 1D and 2D kinematic residual maps and kinematry diagnostic plots for D3a_15504. The orientation is unclear (as indicated by the question mark in Table 3). However, a rough estimate can be deduced from the attenuation, although it could be affected by the presence of the nearby satellite.

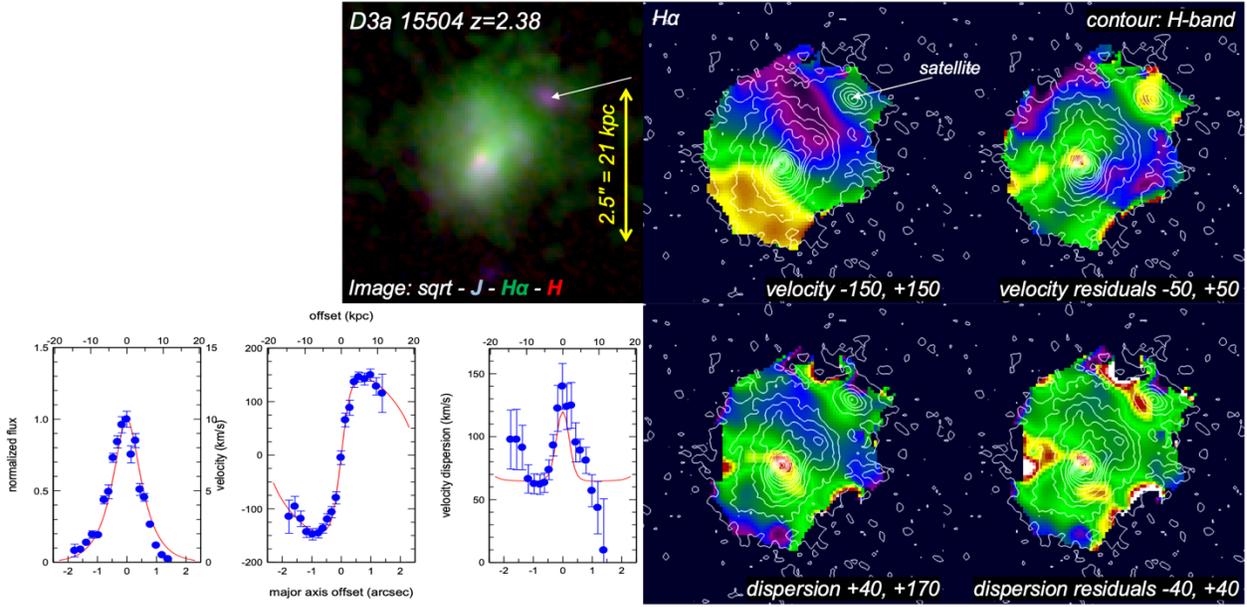

*Figure A14: Top and Right: Images of continuum and Hα velocity, velocity dispersion and their residuals in D3a_15504 (z=2.38). Bottom Left: $1^D$ Hα intensity, velocity and velocity dispersion cuts along $PA_{major}$, as well as the best fit **DYSMAL** model (red line).*

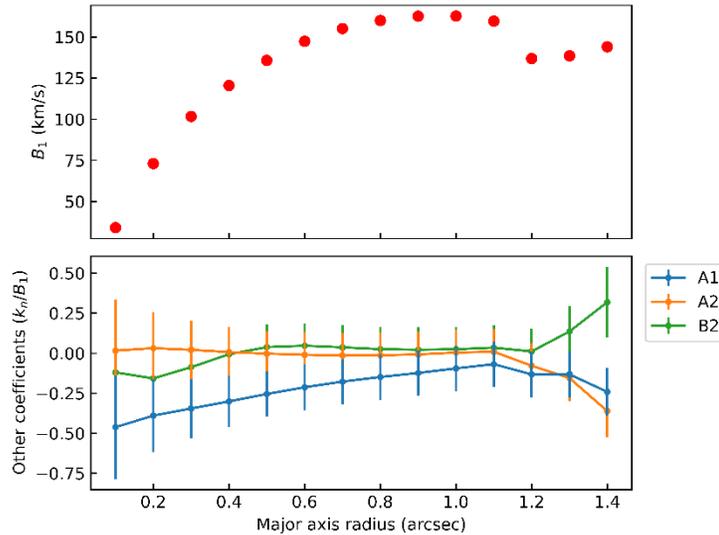

*Figure A15: Kinemetry of Hα velocity maps in D3a_15504.*

## *A.8 D3a_6004 (z=2.39)*

Figures A16 and A17 show the 1D and 2D kinematic residual maps and kinemetry diagnostic plots for D3a_6004. Extinction is used to derive the orientation of the rotation of the source.

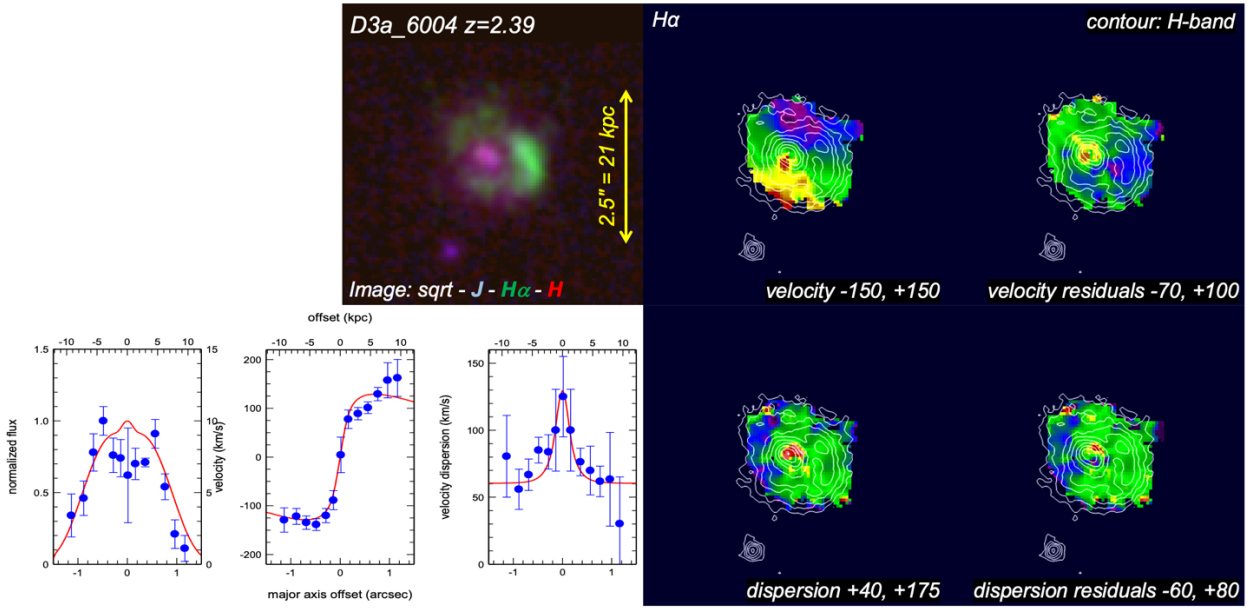

*Figure A16: Top and Right: Images of continuum and Hα velocity, velocity dispersion, and their residuals in D3a_6004 (z=2.39). Bottom Left: 1D Hα intensity, velocity and velocity dispersion cuts along $PA_{major}$, as well as the best-fit **DYSMAL** model (red line).*

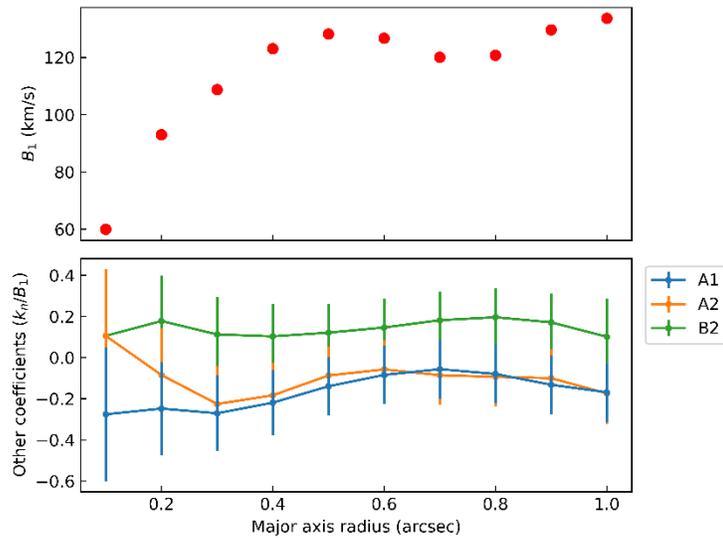

*Figure A17: Kinemetry of Hα velocity maps in D3a_6004.*

# APPENDIX B: NULL HYPOTHESIS TEST USING MOCK DATA

We performed a null hypothesis test to check the significance of our results. To do so we first generated inflow-free data sets based on the kinematic model and noise properties of the BX610 data, using DYSMAL. The angular and spectral resolution were matched to the data. The noise maps were generated using ESSENCE (Tsukui et al. 2023) so that we preserved the correlated noise in the data. We considered the following three cases: (1) similar signal-to-noise ratio (S/N) and angular resolution as the original BX610 data; (2) half the S/N; and (3) same S/N and half the angular resolution.

The mock data was then analyzed in the same way as the analyses presented in the paper. The velocity maps and corresponding residuals for each of the three cases are shown on the upper and lower panels of Figure B1 respectively. In addition, a kinemetry analysis of case 1 is also shown in Figure B2.

As expected, the final mock residual maps do not exhibit the characteristic inflow signature seen in the real data. Likewise, the kinemetry analysis does not indicate any deviation from pure circular rotation. This further confirms the authenticity of the inflow signatures observed in our analyses.

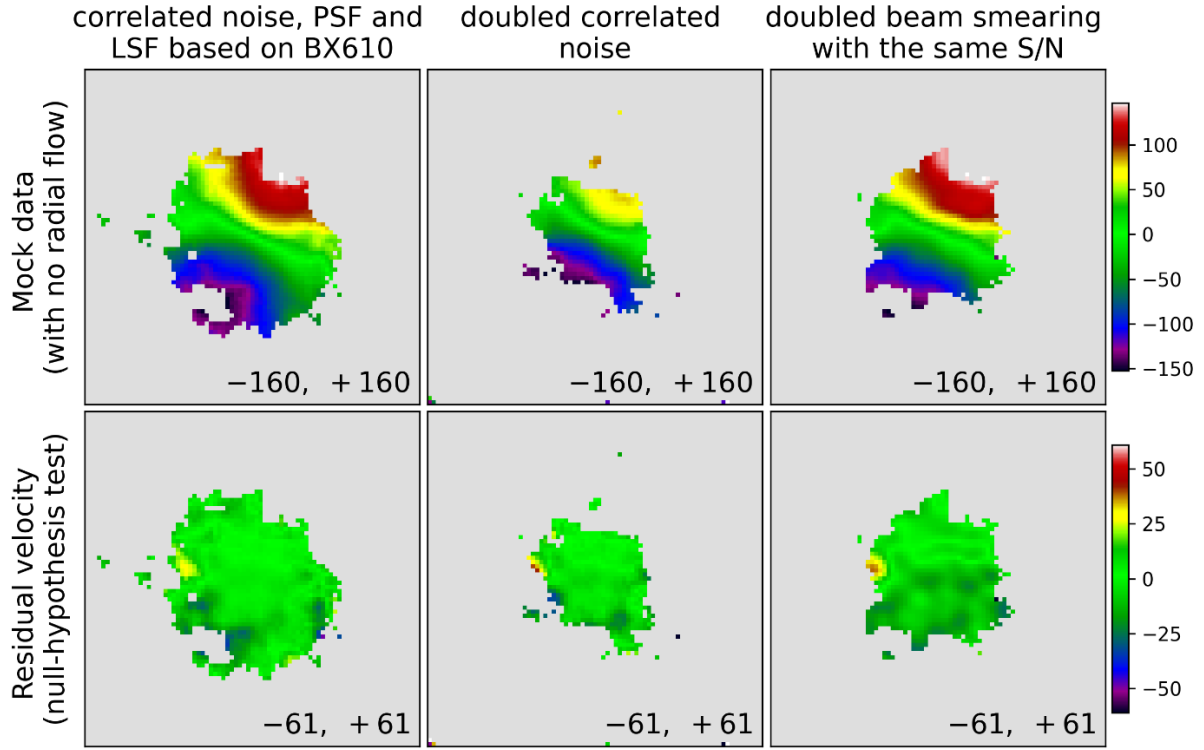

*Figure B1: (Top) Velocity maps of the mock data sets without inflows, for each of the three cases (cases 1– 3, from left to right) and (bottom) corresponding velocity residuals.*

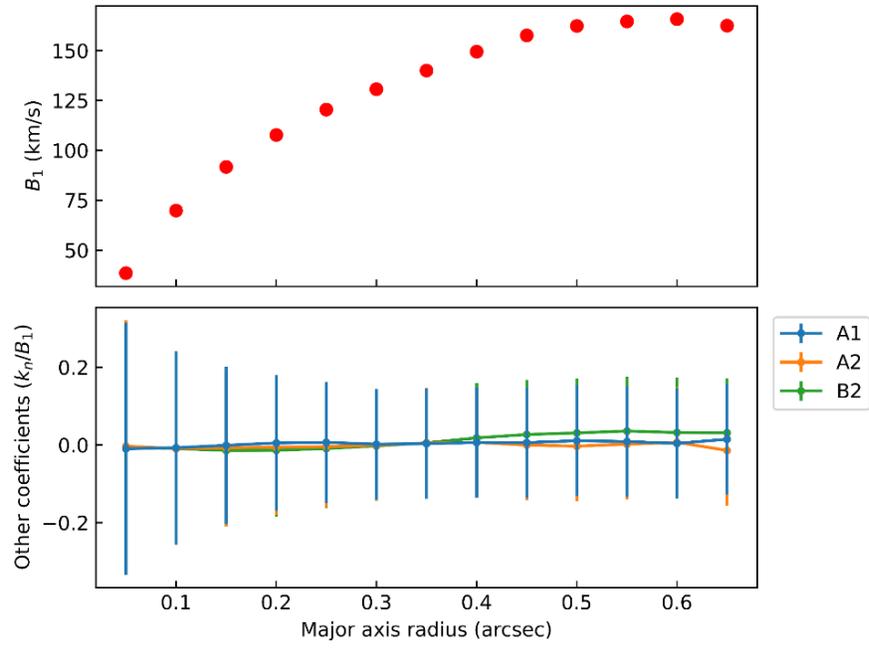

*Figure B2: Kinemetry analysis of case 1.*